\pdfoutput=1
\RequirePackage[T1]{fontenc}
\documentclass[12pt]{article}
\usepackage{lmodern}
\usepackage{amsmath}
\usepackage{amssymb}
\usepackage{amsfonts}
\usepackage{bm}
\usepackage{graphicx}
\usepackage{systeme}
\usepackage{enumerate}
\usepackage{comment}
\usepackage{here}
\usepackage{mathrsfs}
\usepackage{url}
\usepackage{braket}
\usepackage{cite}
\usepackage[svgnames]{xcolor}
\usepackage[colorlinks,citecolor=DarkGreen,linkcolor=FireBrick,urlcolor=FireBrick,breaklinks=true,linktoc=all]{hyperref}
\usepackage{setspace} 
\usepackage[top=30truemm,bottom=30truemm,left=25truemm,right=25truemm]{geometry}

\usepackage{authblk}

\title{Anomaly of Subsystem Symmetries in \\ Exotic and Foliated $BF$ Theories}

\author{Shutaro Shimamura}
\affil{\textit{Department of Physics, The University of Tokyo, Bunkyo-ku, Tokyo 113-0033, Japan}}

\date{}

\renewcommand{\L}{\mathcal{L}}
\newcommand{\Z}{\mathbb{Z}}

\newcommand {\ali}[1] {\begin{align} #1 \end{align}}
\newcommand {\sbali}[1] {\begin{subequations} \begin{align} #1 \end{align} \end{subequations}}
\newcommand {\alis}[1] {\begin{align} \begin{split} #1 \end{split} \end{align}}

\usepackage{xcolor}

\usepackage[status=draft,inline,nomargin]{fixme}
\fxusetheme{color}
\FXRegisterAuthor{ko}{ako}{KO}

\FXRegisterAuthor{shu}{ashu}{\color{magenta}SHU}
    
\begin{document}
\maketitle

\vspace{1.5cm}

\begin{abstract}
    We study the mixed 't Hooft anomaly of the subsystem symmetries in the exotic $BF$ theory and the foliated $BF$ theory in 2+1 dimensions, both of which are fractonic quantum field theories describing the equivalent physics. In the anomaly inflow mechanism, the 't Hooft anomaly of the subsystem symmetries can be canceled by combining a subsystem symmetry-protected topological (SSPT) phase in one dimension higher. In this work, we construct the exotic and foliated $BF$ theories with background gauge fields, and the exotic and foliated forms of the SSPT phases using the foliated-exotic duality. In the foliated form, we see that the non-topological defect that describes a fracton can be viewed as a symmetry-like operator. We also newly construct the foliated and exotic SSPT phases with different foliation structures via the foliated-exotic duality. We can show that the SSPT phases with different foliation structures cancel the same anomaly. This may provide a clue to the characterization of the 't Hooft anomaly of subsystem symmetries.
\end{abstract}

\thispagestyle{empty}
\clearpage
\addtocounter{page}{-1}

\newpage

\setlength{\parskip}{2mm}
\setlength{\abovedisplayskip}{10pt} 
\setlength{\belowdisplayskip}{10pt} 
\numberwithin{equation}{section}

\tableofcontents

\newpage

\section{Introduction}

Certain new types of phases of matter have attracted much attention in recent years. The phases have characteristic excitations that cannot move in space: \textit{fracton}, or can only move along a certain dimensional submanifold like a line: \textit{lineon}, or a plane: \textit{planon}. They are called fracton phases (See for reviews \cite{Nandkishore:2018sel,Pretko:2020cko,Gromov:2022cxa}). Fracton phases were originally motivated by quantum storage and glassy dynamics \cite{Chamon:2004lew,Haah:2011drr,PhysRevLett.111.200501}.

The excitations with characteristic mobility originate from \textit{subsystem global symmetry}, which is one of the generalized global symmetries \cite{Gaiotto:2014kfa}, whose symmetry operator is supported on a submanifold of a certain shape in spacetime. The submanifolds are partially topological; that is, they can only be deformed in certain directions, and the symmetry operators on different submanifolds that cannot be deformed to each other are independent \cite{Seiberg:2019vrp}. Fracton phases have been studied not only as lattice models (earlier studies are in \cite{Paramekanti:2002iup,Haah:2011drr,Vijay:2015mka,Vijay:2016phm}), but also as effective continuum quantum field theories (QFTs) based on tensor gauge fields \cite{Pretko:2018jbi,You:2019ciz,Slagle:2017wrc,Seiberg:2020bhn,Seiberg:2020wsg,Seiberg:2020cxy,Gorantla:2020xap,Gorantla:2021bda,Fontana:2020tby,Ma:2020svo,Gorantla:2020jpy,Yamaguchi:2021qrx,Yamaguchi:2021xeq,Burnell:2021reh,Katsura:2022xkg,Gorantla:2022eem,Gorantla:2022ssr,Honda:2022shd,Luo:2022mrj}. Such fractonic QFTs have discrete rotational symmetry and the tensor gauge fields respect the rotational symmetry. Related to the lattice-like spatial symmetry, fractonic QFTs exhibit UV/IR mixing; the low-energy IR physics depends on some microscopic quantities in short-distance UV physics. As a result of this phenomenon, the tensor gauge fields can have some singularities and discontinuities in certain directions \cite{Seiberg:2020bhn,Seiberg:2020wsg,Seiberg:2020cxy}. Subsystem symmetry in fractonic QFTs is studied in the context of generalized global symmetries, and also related to multipole symmetry \cite{Ebisu:2023idd,Ebisu:2024cke}, non-invertible symmetry \cite{Cao:2023doz} and Symmetry TFT \cite{Cao:2023rrb}.

Although tensor gauge fields can describe some fracton phases, the QFTs constructed from \textit{foliation} structure also exhibit fractonic features, and such QFTs are called foliated QFTs \cite{Slagle:2018swq,Slagle:2020ugk,Hsin:2021mjn,Geng:2021cmq,Ohmori:2022rzz,Spieler:2023wkz,Cao:2023rrb,Ebisu:2023idd,Ebisu:2024cke,Hsin:2023ooo,Hsin:2024eyg}.\footnote{Some lattice models can also be constructed by using foliation structure. They describe the foliated fracton phases \cite{Shirley:2017suz,Shirley:2018nhn,Shirley:2018vtc,Shirley:2019uou,Shirley:2022wri}.} A foliation is a decomposition of a manifold into an infinite number of submanifolds: \textit{leaves}. In the foliated QFT context, we consider foliations where the intersections of leaves form a lattice-like structure. Foliated QFTs contain foliated gauge fields, which are considered as lower form gauge fields on the leaves, and bulk gauge fields, which mediate the foliated gauge fields. In contrast to foliated QFTs, we call fractonic QFTs with tensor gauge fields exotic QFTs. Some foliated QFTs are equivalent to the corresponding exotic QFTs \cite{Hsin:2021mjn,Ohmori:2022rzz,Spieler:2023wkz,Cao:2023rrb,Ebisu:2023idd}: this is called the foliated-exotic duality. For example, the low-energy theory of the X-cube model \cite{Vijay:2016phm} is described by both of the exotic and foliated $BF$ theory in 3+1 dimensions \cite{Seiberg:2020cxy,Slagle:2020ugk}, and the tensor gauge fields in the exotic $BF$ theory explicitly correspond to the foliated and bulk gauge fields in the foliated $BF$ theory \cite{Ohmori:2022rzz}.\footnote{These fractonic $BF$ theories are similar to the ordinary relativistic $BF$ theory \cite{Maldacena:2001ss,Banks:2010zn,Kapustin:2014gua} and have several analogies with it.} This structure is specific to fracton phases, and is considered as a new type of duality.

On another topic, consider a $d$-dimensional QFT with a global symmetry $G$. The global symmetry acts on a charged object as a global transformation. Then, we can couple the global symmetry to a background gauge field $A$ for $G$, and replace the parameter of the global transformation with a local parameter, which is absorbed in the gauge transformation of $A$. In this situation, the partition function may not be invariant under the local transformation $A \rightarrow A^g$:
\ali{
    Z[A^g] = e^{i\alpha(A,g)} Z[A] \,.
}
If the function $\alpha(A,g)$ cannot be canceled by a local counterterm of $A$ and $g$, the QFT is said to have an 't Hooft anomaly \cite{tHooft:1979rat}. Due to the non-invariance of the partition function, we cannot sum over the background gauge field $A$; the global symmetry cannot be gauged. In the anomaly inflow mechanism \cite{Callan:1984sa}, an 't Hooft anomaly in a $d$-dimensional QFT is captured by a classical field theory in $d+1$ dimensions. In relativistic QFTs, the classical field theories are called invertible field theories \cite{Freed:2016rqq} or symmetry-protected topological 
 (SPT) phases \cite{Gu:2009dr,Chen:2010gda}. In this case, the classification of anomalies is interpreted as the cobordism group classification of SPT phases\cite{Chen:2011pg,Wen:2013oza,Kapustin:2014tfa,Kapustin:2014dxa}.

Then, what about the case of subsystem symmetry? In some exotic theories \cite{Burnell:2021reh,Luo:2022mrj} and simple foliated theories \cite{Hsin:2021mjn}, anomalies of the subsystem symmetries are captured by classical field theories called subsystem symmetry-protected topological (SSPT) phases \cite{You:2018oai,Devakul:2018fhz}. However, the relation between 't Hooft anomalies of subsystem symmetry and SSPT phases is obscure. For example, the foliation structure of the bulk SSPT phase is not canonically determined \cite{Burnell:2021reh,Luo:2022mrj}. 
This fact obstructs the classification of anomalies of subsystem symmetry.

In this paper, we discuss the $\Z_N \times \Z_N$ mixed 't Hooft anomaly\footnote{If two global symmetries can be gauged respectively, but cannot be gauged simultaneously, the system is said to have a mixed 't Hooft anomaly.} of subsystem symmetries in the exotic and foliated $BF$ theory in 2+1 dimensions \cite{Seiberg:2020bhn,Ohmori:2022rzz}. Although they are equivalent, it is easier to couple the exotic $BF$ theory to background tensor gauge fields and construct the 3+1d exotic SSPT phase with two simultaneous foliations that cancels the 't Hooft anomaly of it. After constructing them, we will assume field correspondences between the exotic and foliated $BF$ theories extending the previous result without background gauge fields \cite{Ohmori:2022rzz}, and construct the 2+1d foliated $BF$ theory coupled to background foliated and bulk gauge fields. In the foliated $BF$ theory, we find that the non-topological defect that describes a fracton is considered as a symmetry-like operator. Next, using the field correspondences, we will construct the foliated form of the 3+1d SSPT phase with two foliations that cancels the 't Hooft anomaly of the foliated $BF$ theory. Finally, we will construct a bulk SSPT phase with three simultaneous foliations. In the foliated form, it is simple to construct the SSPT phase with three foliations from the SSPT phase with two foliations we have constructed. In the exotic form, on the other hand, the relation between the two SSPT phases is not manifest. Here we will construct the exotic form of the SSPT phase with three foliations via the foliated-exotic duality. This can be seen as a systematic way to construct exotic QFTs with different foliation structures. In addition, we will see that these two foliated SSPT phases are connected via a smooth deformation. This fact is considered as a hint for characterizing 't Hooft anomalies of subsystem symmetry.

\begin{center}
{\subsection*{\textmd{\textit{Organization}}}}
\end{center}

The organization of the rest of the paper is as follows. 

In Section \ref{section 2}, we will consider the anomaly of the 2+1d exotic $BF$ theory. In Section \ref{section 21}, we review the the 2+1d exotic $BF$ theory and its subsystem symmetries \cite{Seiberg:2020bhn,Ohmori:2022rzz}. In Section \ref{section 22}, we consider the 2+1d exotic $BF$ theory coupled to background tensor gauge fields. In Section \ref{section 23}, we construct the exotic form of the 3+1d SSPT phase with two foliations that cancels the anomaly of the 2+1d exotic $BF$ theory. 

In Section \ref{section 3}, we will consider the anomaly of the 2+1d foliated $BF$ theory \cite{Ohmori:2022rzz}. In Section \ref{section 31}, we review the 2+1d foliated $BF$ theory and the foliated-exotic duality in the foliated and exotic $BF$ theories. In Section \ref{section 32}, we expand the field correspondences to the background gauge fields and construct the 2+1d foliated $BF$ theory coupled to background foliated and bulk gauge fields. In Section \ref{section 33}, we convert the SSPT phase with two foliations from the exotic form into the foliated form.

In Section \ref{section 4}, we will discuss changing the foliation structure from two foliations to three foliations. The SSPT phase with three foliations also cancels the same anomaly of the 2+1d exotic/foliated $BF$ theory. In Section \ref{section 41}, we see the change is easily carried out in the foliated SSPT phase. In Section \ref{section 42}, we convert the SSPT phase with three foliations from the foliated form into the exotic form by assuming field correspondences.

\vskip\baselineskip

\section{Anomaly in the 2+1d Exotic $BF$ Theory} \label{section 2}

In this section, we review the exotic $BF$ theory in 2+1 dimensions \cite{Seiberg:2020bhn,Ohmori:2022rzz}, which is the low-energy effective QFT of the $\Z_N$ plaquette Ising model \cite{Johnston:2016mbz}, and consider the two types of subsystem symmetries and their mixed 't Hooft anomaly. Due to the anomaly, we cannot gauge both of the subsystem symmetries simultaneously. This anomaly can be canceled by a classical field theory in one dimension higher, which is called a subsystem symmetry-protected topological (SSPT)
phase\cite{You:2018oai,Devakul:2018fhz}. We will consider the mixed 't Hooft anomaly and the SSPT phase associated with the 2+1d exotic $BF$ theory. In the literature \cite{Burnell:2021reh}, they studied the mixed 't Hooft anomaly and the SSPT phase associated with the 3+1d exotic $BF$ theory, and we basically proceed in parallel with that.

\subsection{Exotic $BF$ Theory and Symmetries} \label{section 21}

We take a three-torus of lengths $l^0$, $l^1$, $l^2$ as a Euclidean spacetime and the coordinates $(x^0,x^1,x^2)$ on it. We consider the exotic $BF$ theory, whose rotational symmetry is only the $90$ degree ones. Such a theory has tensor gauge fields, each of which is in a representation of the $90$ degree rotation group $\Z_4$. Irreducible representations of $\mathbb{Z}_4$ are one-dimensional ones $\bm{1}_n\, (n=0, \pm 1, 2)$, where $n$ is the spin. The 2+1d exotic $BF$ theory contains a compact scalar $\hat{\phi}^{12}$ in the representation $\bm{1}_2$ and a $U(1)$ tensor gauge fields $(A_0,A_{12})$ in the representations $(\bm{1}_0,\bm{1}_2)$. Their gauge transformations are
\ali{
     \hat{\phi}^{12} &\rightarrow \hat{\phi}^{12} + 2\pi \hat{m}^1 - 2\pi \hat{m}^2 \,, \label{egauge3} \\
    A_{0} &\rightarrow A_{0} + \partial_0 \alpha \,, \label{egauge1} \\
    A_{12} &\rightarrow A_{12} + \partial_1\partial_2 \alpha \,, \label{egauge2} 
}
where $\hat{m}^k$ is an $x^k$-dependent integer-valued gauge parameter, and $\alpha$ is a gauge parameter in the representation $\bm{1}_0$. The gauge parameter $\alpha$ has its own gauge transformation: $\alpha \rightarrow \alpha + 2\pi n^1 + 2\pi n^2$, where $n^k$ is an $x^k$-dependent integer-valued gauge parameter. Due to constant parts of $\hat{m}^k$ and $n^k$, $\hat{\phi}^{12}$ and $\alpha$ can be regarded as $U(1)$-valued: $\hat{\phi}^{12} \sim \hat{\phi}^{12} + 2\pi$, $\alpha \sim \alpha + 2\pi$. These tensor gauge fields and parameters can have particular types of singularities and discontinuities \cite{Seiberg:2020bhn,Ohmori:2022rzz}.

The exotic $BF$ Lagrangian is\footnote{The subscript e means that the theory is written by using tensor gauge fields, which we call the exotic form. Also, the subscript f, which will appear later, means that the theory is written by using foliated gauge fields and bulk gauge fields, which we call the foliated form.}
\ali{
    \L_{\text{e}} = \frac{i N}{2\pi}\hat{\phi}^{12}(\partial_0 A_{12} - \partial_{1}\partial_{2} A_0) \,. \label{elagrangian}
}

The equations of motion are
\ali{
    &\frac{N}{2\pi} \partial_{1}\partial_{2} \hat{\phi}^{12} = 0 \,, \label{eeom1}\\
    &\frac{N}{2\pi} \partial_{0} \hat{\phi}^{12} = 0 \,, \label{eeom2} \\
    &\frac{N}{2\pi} (\partial_0 A_{12} - \partial_{1}\partial_{2} A_0) = 0 \,. \label{eeom3}
}

Let us discuss symmetries. The subsystem symmetries are described by the partially topological gauge-invariant defects and operators. Since the fractonic theory is not fully rotational invariant, the time and space directions must be treated in different manners even in Euclidean spacetime. This fact implies that we have two types of symmetries: space-like symmetry and time-like symmetry \cite{Gorantla:2022eem}. A space-like symmetry has a charged operator in space, and the corresponding symmetry operator acts on the charged operator. On the other hand, a time-like symmetry has a time-like charged defect, whose manifold is a trajectory of an infinitely heavy particle, and the corresponding symmetry operator in space can remotely detect the time-like defect.\footnote{In relativistic Euclidean QFT, which has the full rotational spacetime symmetry, time-like symmetry and space-like symmetry are not distinguished.}

The exotic $BF$ theory has two types of space-like subsystem symmetries. One is the $\Z_N$ electric global symmetry that is generated by the symmetry operator
\begin{align}
    \tilde{V}[x] = \exp \left[ i \hat{\phi}^{12} \right]\,. \label{epoint}
\end{align}
The charged operators are the strip operators
\begin{align}
    \tilde{W}_1\left[ S^1_2 \right] = \exp \left[ i  \oint_{S^1_2}  dx^2 dx^1 A_{12} \right] \,, \label{sestrip1}\\
    \tilde{W}_2\left[ S^2_2  \right] = \exp \left[ i \oint_{S^2_2}  dx^1 dx^2 A_{12}  \right]\,, \label{sestrip2}
\end{align}
where $S^k_2$ is a two-dimensional strip with a fixed width along the $x^k$ direction in the $(x^1,x^2)$-plane. $\tilde{V}[x]$ acts on $\tilde{W}_k\left[ S^k_2 \right]$ as
\begin{align}
    \tilde{V}[x]  \tilde{W}_k\left[S^k_2\right] \tilde{V}[x]^{-1} =  e^{2\pi i/N} \  \tilde{W}_k \left[S^k_2 \right]  \ \ ,\quad \text{if} \quad x^k_1 < x^k < x^k_2\,, \label{eactvw}
\end{align}
where $S^k_2$ has the width of $[x^k_1, x^k_2]$. The $\Z_N$ electric global symmetry is a space-like subsystem symmetry on a zero-dimensional submanifold. For the action on the field, $\tilde{V}[x]$ acts as
\ali{
    A_{12} \rightarrow A_{12} + \Lambda_{12} \,,
}
where $\Lambda_{12}$ is a flat $\Z_N$ field, which satisfies
\ali{
     \oint_{S^k_2}  dx^j dx^k  \Lambda_{12} \in \frac{2\pi}{N} \Z \,,\quad  ((k,j) = (1,2), (2,1))\,,
}
so then $\Lambda_{12}$ can be written as $\Lambda_{12} = \frac{2\pi}{N}\left( \frac{1}{l^2} \partial_1 n_1(x^1) + \frac{1}{l^1} \partial_2 n_2(x^2)  \right) $, where $n_k(x^k)$ is an $x^k$-dependent integer-valued and single-valued function.

The other space-like symmetries are the $\Z_N$ dipole global symmetries that are generated by the strip operators $\tilde{W}_k[ S^k_2 ]$ in \eqref{sestrip1} and \eqref{sestrip2}. The charged operator is $\tilde{V}[x]$ in \eqref{epoint}, and the actions are
\ali{
    \tilde{W}_k\left[S^k_2\right] \tilde{V}[x] \tilde{W}_k\left[S^k_2\right]^{-1} =  e^{-2\pi i/N} \  \tilde{V}[x]  \ \ ,\quad \text{if} \quad x^k_1 < x^k < x^k_2\,. \label{eactwv}
}
For the action on the field, $\tilde{W}_k\left[ S^k_2 \right]$ acts as
\ali{
    \hat{\phi}^{12} \rightarrow \hat{\phi}^{12} + \hat{\Lambda}^{12} \,,
}
where $\hat{\Lambda}^{12}$ is valued in $2\pi \Z/N$, so $\hat{\Lambda}^{12}$ can be written as  $\hat{\Lambda}^{12} = \frac{2\pi}{N}\left(n_1(x^1)  + n_2(x^2) \right)$, where $n_k(x^k)$ is an $x^k$-dependent integer-valued and single-valued function.

From the actions \eqref{eactvw} and \eqref{eactwv}, each symmetry operator is the charged operator of the other, which is similar to the symmetries in the ordinary $BF$ theory \cite{Maldacena:2001ss,Banks:2010zn,Kapustin:2014gua}. From this fact, we expect that the two types of subsystem symmetries have a mixed 't Hooft anomaly.

We also have a time-like symmetry whose charged defect describes a fracton, which is called the $\Z_N$ tensor symmetry. The symmetry operator is a quadrupole operator:
\begin{align}
    \tilde{T}\left[C_1^{12,\text{rect}}(x^1_1,x^1_2,x^2_1,x^2_2)\right] = \exp\left[-i \Delta_{12} \hat{\phi}^{12} (x^1_1,x^1_2,x^2_1,x^2_2) \label{etimelikeop}   \right] \,,
\end{align}
where $\Delta_{12} \hat{\phi}^{12}(x^1_1,x^1_2,x^2_1,x^2_2) = \hat{\phi}^{12}(x^1_2,x^2_2) - \hat{\phi}^{12}(x^1_2,x^2_1) - \hat{\phi}^{12}(x^1_1,x^2_2) + \hat{\phi}^{12}(x^1_1,x^2_1)$, and \\
\noindent  $C_1^{12,\text{rect}}(x^1_1,x^1_2,x^2_1,x^2_2)$ is a rectangle whose vertices are the four points above. This operator is a product of the operators $\tilde{V} = \exp \left[ i \hat{\phi}^{12} \right]$ localized at the corners of the rectangle. The charged defect is
\begin{align}
    \tilde{F}[C_1^0] = \exp \left[ i \oint_{C_1^0} dx^0 A_0\right] \,, \label{efracton}
\end{align}
where $C_1^0$ is a closed one-dimensional loop along the time $x^0$ direction. The deformation of $C^0_1$ would break the gauge invariance of the defect, which means that this defect describes a fracton that cannot move alone in space. The operator $\tilde{T}\left[C_1^{12,\text{rect}}(x^1_1,x^1_2,x^2_1,x^2_2)\right]$ detects the fracton defect $\tilde{F}[C_1^0]$ as
\begin{align}
    \tilde{T}\left[C_1^{12,\text{rect}}(x^1_1,x^1_2,x^2_1,x^2_2)\right] \cdot \tilde{F}[C^0_1] = e^{-2\pi i/N} \tilde{F}[C^0_1] \,, \label{tfctvw}
\end{align}
when $C_1^{12,\text{rect}}(x^1_1,x^1_2,x^2_1,x^2_2)$ surrounds $C^0_1$.\footnote{The edges of a rectangle $C_1^{12,\text{rect}}(x^1_1,x^1_2,x^2_1,x^2_2)$ cannot be remotely detected by other operators, but the operator $\tilde{T}[C_1^{12,\text{rect}}(x^1_1,x^1_2,x^2_1,x^2_2)]$ is actually an operator on a rectangle. In fractonic theory, operators are not necessarily remotely detectable unlike ordinary topological order or topological field theory \cite{Ohmori:2022rzz}} For the action on the field, $\tilde{T}\left[C_1^{12,\text{rect}}(x^1_1,x^1_2,x^2_1,x^2_2)\right]$ acts as
\ali{
    A_0 \rightarrow A_0 + \Lambda_0 \,,
}
where $\Lambda_0$ is a $\Z_N$ field, which satisfies
\ali{
     \oint_{C_1^0} dx^0  \Lambda_0 \in \frac{2\pi}{N} \Z \,,
}
so then $\Lambda_0$ can be written as $ \Lambda_0 = \frac{2\pi}{N} \frac{1}{l^0} ( n_1(x^1)  + n_2(x^2))$, where $n_k(x^k)$ is an $x^k$-dependent integer-valued and single-valued function\cite{Gorantla:2022eem}.

In addition, we can construct the defect that describes a dipole of fractons separated in the $x^k$ direction. A dipole of fractons can be represented as
\ali{
    \tilde{F}[C_1^0(x^k_1,x^j)]  \tilde{F}[C_1^0(x^k_2,x^j)]^{-1}  = \exp \left[ i \oint_{C_1^0}\int^{x^k_2}_{x^k_1} dx^0 dx^k \partial_k A_0\right] \,,
}
where $C_1^0(x^1,x^2)$ is a closed loop along the $x^0$ direction at a point $(x^1,x^2)$ in space, and $(k,j)=(1,2),(2,1)$. This defect is partially topological, that is, it can be deformed to the strip defect
\begin{align}
    \tilde{W}_{k,\text{dip}} \left[ S^{k,\text{dip}}_2 \right] &= \exp \left[ i \oint_{S^{k,\text{dip}}_2} ( dx^0 dx^k \partial_k A_0 + dx^j dx^k A_{12} )  \right] \,, \label{estripk}
\end{align}
where $S^{k,\text{dip}}_2$ is now a two-dimensional strip with a fixed width along the $x^k$ direction in spacetime.

\subsection{Coupling to the Background Tensor Gauge Fields} \label{section 22}

In this section, we couple the subsystem symmetries to background gauge fields and replace the parameters of the symmetry actions on the fields with local transformations. Then, the local transformations are absorbed into the gauge transformations of the background gauge fields. To gauge these symmetries, one has to sum over configurations of the gauge fields. If the partition function is not invariant under the gauge transformations, we cannot gauge all the symmetries at the same time, indicating a mixed 't Hooft anomaly. We will see that the partition function of the 2+1d exotic $BF$ theory is indeed not gauge invariant and the subsystem symmetries have a mixed 't Hooft anomaly.

The tensor time-like symmetry $\tilde{T}\left[C_1^{12,\text{rect}}(x^1_1,x^1_2,x^2_1,x^2_2)\right]$ and the electric space-like symmetry $\tilde{V}[x]$ are coupled to the $U(1)$ gauge field $C_{012}$ in $\bm{1}_2$. The dipole symmetries $\tilde{W}_k$ are coupled to the $U(1)$ gauge fields $(\hat{C}^{12}_0, \hat{C})$ in $(\bm{1}_2, \bm{1}_0)$. We will see later that the background gauge transformations of $C_{012}$ and $(\hat{C}^{12}_0, \hat{C})$ are absorbed into the local transformations of $(A_0, A_{12})$ and $\hat{\phi}^{12}$. The tensor gauge fields $(A_0, A_{12})$ transform as
\begin{align}
    A_0 &\rightarrow A_0 + \Lambda_0 \,, \label{a0bk} \\
    A_{12} &\rightarrow A_{12} + \Lambda_{12} \,, \label{a12bk}
\end{align}
where $\Lambda_0$ and $\Lambda_{12}$ are background gauge parameters. Then, the background gauge transformations of $C_{012}$ is
\begin{align}
    C_{012} \rightarrow C_{012} +  \partial_0 \Lambda_{12} - \partial_1 \partial_2 \Lambda_0 \,.
\end{align}
The tensor gauge field $\hat{\phi}^{12}$ transforms as
\begin{align}
    \hat{\phi}^{12} \rightarrow \hat{\phi}^{12} +  \hat{\Lambda}^{12} \,, \label{phi12bk}
\end{align}
where $\hat{\Lambda}^{12}$ is a background gauge parameter. Then, the background gauge transformations of $(\hat{C}^{12}_0, \hat{C})$ are
\begin{align}
    \hat{C}^{12}_0 &\rightarrow \hat{C}^{12}_0 + \partial_0 \hat{\Lambda}^{12} \,, \\
    \hat{C} &\rightarrow \hat{C} + \partial_1 \partial_2 \hat{\Lambda}^{12} \,.
\end{align}

The Lagrangian including the background gauge fields is
\begin{align}
\begin{split}
    \L_{\text{e}} \left[ C_{012}, \hat{C}_0^{12}, \hat{C} \right] = &\frac{iN}{2\pi} \left[ \hat{\phi}^{12} (\partial_0 A_{12} - \partial_1 \partial_2 A_0 -C_{012} ) + A_{12} \hat{C}^{12}_0 + A_0 \hat{C} \right] \\ 
    &+ \frac{iN}{2\pi} \chi (\partial_0 \hat{C} - \partial_1 \partial_2 \hat{C}^{12}_0) + \frac{iN}{2\pi}\hat{\chi}^{12} C_{012} \,, \label{ebfc}
\end{split}
\end{align}
Since the symmetries coupled to the $U(1)$ gauge fields are $\Z_N$ symmetries, we need terms of $\chi$ and $\hat{\chi}^{12}$ that are dynamical fields, so that the $U(1)$ gauge fields $C_{012}$ and $(\hat{C}^{12}_0, \hat{C})$ are restricted to $\Z_N$ gauge fields. $\chi$ is $U(1)$-valued, and $\hat{\chi}^{12}$ is $2\pi \Z$-valued. Their dynamical gauge transformations are
\begin{align}
    \chi &\rightarrow \chi + \alpha \,, \\
    \hat{\chi}^{12} &\rightarrow \hat{\chi}^{12} + 2\pi \hat{m}^1 - 2\pi \hat{m}^2 \,.
\end{align}

Under the background gauge transformations, the Lagrangian transforms as
\begin{align}
\begin{split}
     \delta_g \L_{\text{e}} \left[ C_{012}, \hat{C}_0^{12}, \hat{C} \right] &=  \frac{iN}{2\pi}  \left[  \hat{\Lambda}^{12} (\partial_0 A_{12} - \partial_1 \partial_2 A_0 -C_{012} ) + \Lambda_{12} (\hat{C}^{12}_0 + \partial_0 \hat{\Lambda}^{12}) \right. \\
    & \quad \quad \quad \left. + \Lambda_0 (\hat{C} + \partial_1 \partial_2 \hat{\Lambda}^{12}) + A_{12} \partial_0 \hat{\Lambda}^{12} + A_0 \partial_1 \partial_2 \hat{\Lambda}^{12}\right] \\
    &= \frac{iN}{2\pi}  \left[  -\hat{\Lambda}^{12} C_{012}  + \Lambda_{12} (\hat{C}^{12}_0 + \partial_0 \hat{\Lambda}^{12})  + \Lambda_0 (\hat{C} + \partial_1 \partial_2 \hat{\Lambda}^{12}) \right] \,. \label{eanomaly}
\end{split}
\end{align}
If $C_{012}$ and $(\Lambda_0, \lambda_{12})$ are absent or $(\hat{C}^{12}_0, \hat{C})$ and $\hat{\Lambda}^{12}$ are absent, the partition function is invariant. So we can gauge the one side of the symmetry solely, but we cannot gauge both of the symmetries simultaneously. It is a mixed 't Hooft anomaly for subsystem symmetries $\Z_N \times \Z_N$  \cite{Burnell:2021reh}.

\subsection{Exotic SSPT Phase in 3+1 Dimensions} \label{section 23}

We saw the exotic BF theory in 2+1 dimensions has the mixed 't Hooft anomaly. This anomaly can be canceled by a classical field theory in one dimension higher, which is the continuum  description of what is called a SSPT phase. We consider the SSPT phase in 3+1 dimensions with the coordinates $(x^0,x^1,x^2,x^3)$ (the range of $x^3$ will be mentioned later). The foliation structure \cite{Slagle:2018swq,Slagle:2020ugk,Hsin:2021mjn} of the SSPT phase is $x^1$ and $x^2$ foliations, so it is a fractonic system with two simultaneous foliations.\footnote{The foliation structure is similar to the lattice. We will explain foliation in Section \ref{section 31}. For example, $x^1$ foliation on a three dimensional space is a decomposition into an infinite number of planes orthogonal to the $x^1$ direction, so the 
space has lattice-like structure in the $x^1$ direction.} The SSPT phase has the 90 degree discrete rotational symmetry $\Z_4$ for $(x^1,x^2)$ and the continuous rotational symmetry $SO(2)$ for $(x^0,x^1)$ as the spacetime rotational symmetry. This theory has the background gauge fields $(C_{012}, C_{312},C_{[03]})$ and $(\hat{C}^{12}_0, \hat{C}, \hat{C}^{12}_3)$, which are representations of $\Z_4 \times SO(2)$. $C_{012}$, $C_{312}$, $\hat{C}^{12}_0$ and $\hat{C}^{12}_3$ are in $\bm{1}_2$ of $\Z_4$,  $\hat{C}$ is in $\bm{1}_0$ of $\Z_4$ and $C_{[03]}$ is an anti-symmetric tensor of $SO(2)$. To restrict these fields to $\Z_N$, we introduce dynamical gauge fields $\hat{\beta}^{12}$ in $\bm{1}_2$ and $(\beta_0, \beta_{12}, \beta_3)$ in $(\bm{1}_0, \bm{1}_2, \bm{1}_0)$. The background gauge transformations of $(C_{012}, C_{312},C_{[03]})$ are
\begin{align}
    C_{012} &\rightarrow C_{012} + \partial_0 \Lambda_{12} -\partial_1\partial_2 \Lambda_0 \,, \\
    C_{312} &\rightarrow C_{312} + \partial_3 \Lambda_{12} -\partial_1\partial_2 \Lambda_3 \,, \\
    C_{[03]} &\rightarrow C_{[03]} + \partial_0 \Lambda_3 -\partial_3 \Lambda_0 \,.
\end{align}
The background gauge transformations of $(\hat{C}^{12}_0, \hat{C}, \hat{C}^{12}_3)$ are
\begin{align}
    \hat{C}^{12}_0 &\rightarrow \hat{C}^{12}_0 + \partial_0  \hat{\Lambda}^{12} \,, \\
    \hat{C} &\rightarrow \hat{C} + \partial_1 \partial_2  \hat{\Lambda}^{12} \,, \\
    \hat{C}^{12}_3 &\rightarrow \hat{C}^{12}_3 + \partial_3  \hat{\Lambda}^{12} \,,
\end{align}
The background gauge transformations of $\hat{\beta}^{12}$ and $(\beta_0, \beta_{12}, \beta_3)$ are
\begin{align}
    \hat{\beta}^{12} &\rightarrow \hat{\beta}^{12} + \hat{\Lambda}^{12} \,, \\
    \beta_0 &\rightarrow \beta_0 - \Lambda_0 \,, \\
    \beta_{12} &\rightarrow \beta_{12} - \Lambda_{12} \,, \\
    \beta_3 &\rightarrow \beta_3 - \Lambda_3 \,.
\end{align}
In addition, $\hat{\beta}^{12}$ and $(\beta_0, \beta_{12}, \beta_3)$ have dynamical gauge transformations:
\ali{
    \hat{\beta}^{12} &\rightarrow \hat{\beta}^{12} + 2\pi \hat{s}^1 - 2\pi \hat{s}^2 \,, \\
    \beta_0 &\rightarrow \beta_0 + \partial_0 s \,, \\
    \beta_{12} &\rightarrow \beta_{12} + \partial_1 \partial_2 s \,, \\
    \beta_3 &\rightarrow \beta_3 + \partial_3 s \,,
}
where $\hat{s}^k$ is an $x^k$-dependent integer-valued gauge parameter, and $s$ is a gauge parameter in $\bm{1}_0$. 

The SSPT phase is described by the Lagrangian
\begin{align}
\begin{split}
    &\L_{\text{SSPT,e}}\left[ C_{012}, C_{312},C_{[03]}, \hat{C}^{12}_0, \hat{C}, \hat{C}^{12}_3 \right] \\
    & \quad =  \frac{iN}{2\pi} \hat{\beta}^{12} \left( \partial_3 C_{012} - \partial_0 C_{312} - \partial_1 \partial_2 C_{[03]} \right) \\
    & \quad \quad +  \frac{iN}{2\pi} \left[ \beta_0 \left( \partial_3 \hat{C} - \partial_1 \partial_2 \hat{C}^{12}_3  \right) + \beta_{12} \left( \partial_3 \hat{C}^{12}_0 - \partial_0 \hat{C}^{12}_3  \right)  - \beta_3 \left( \partial_0 \hat{C} - \partial_1 \partial_2 \hat{C}^{12}_0  \right) \right] \\
    &\quad \quad  + \frac{iN}{2\pi}  \left( C_{012} \hat{C}^{12}_3 - C_{312} \hat{C}^{12}_0 + C_{[03]} \hat{C}  \right) \,. \label{sspte}
\end{split}
\end{align}
If the theory is on spacetime without a boundary, it is gauge invariant. However if spacetime has a boundary, the partition function of the 3+1d SSPT phase is not invariant. From the anomaly inflow mechanism \cite{Callan:1984sa}, this variation is expected to be canceled by the anomaly of the 2+1d exotic $BF$ theory on the boundary \eqref{eanomaly}. To see this, we put the SSPT phase on the region $x^3 \geq 0$ with the boundary $x^3 = 0$. From the gauge invariance,  the boundary conditions of $\hat{\beta}^{12}$, $\beta_0$ and $\beta_{12}$ are
\ali{
    \hat{\beta}^{12}\, |_{x^3 = 0} &= 0 \,, \\ 
    \beta_0\, |_{x^3 = 0} &= 0 \,, \\
    \beta_{12}\, |_{x^3 = 0} &= 0 \,.
}
On the boundary, we put the 2+1d exotic $BF$ theory coupled to the background gauge fields \eqref{ebfc}, and the background gauge fields in the 3+1d SSPT phase are related to those in the 2+1d exotic $BF$ theory as
\ali{
    C_{\text{SSPT},012}\, |_{x^3 = 0} &= C_{BF,012} \,, \\
    \hat{C}^{12}_{\text{SSPT},0}\, |_{x^3 = 0} &= \hat{C}^{12}_{BF,0} \,, \\ 
    \hat{C}_{\text{SSPT}}\, |_{x^3 = 0} &= \hat{C}_{BF} \,.
}
Note that while the background gauge fields in the 3+1d SSPT phase are restricted to $\Z_N$ tensor gauge fields by the dynamical fields $\hat{\beta}^{12}$ and $(\beta_0, \beta_{12}, \beta_3)$, those in the 2+1d exotic $BF$ theory are restricted by the dynamical fields $\chi$ and $\hat{\chi}^{12}$.
Then, under the background gauge transformations, the Lagrangian transforms as
\begin{align}
\begin{split}
    \delta_g \L_{\text{SSPT,e}} &=  \frac{iN}{2\pi} \hat{\Lambda}^{12} \left( \partial_3 C_{012} - \partial_0 C_{312} - \partial_1 \partial_2 C_{[03]} \right) \\
    & \quad +  \frac{iN}{2\pi} \left[ -\Lambda_0 \left( \partial_3 \hat{C} - \partial_1 \partial_2 \hat{C}^{12}_3  \right) - \Lambda_{12} \left( \partial_3 \hat{C}^{12}_0 - \partial_0 \hat{C}^{12}_3  \right)  + \Lambda_3 \left( \partial_0 \hat{C} - \partial_1 \partial_2 \hat{C}^{12}_0  \right) \right] \\
    &\quad  + \frac{iN}{2\pi}  \left( C_{012} \partial_3 \hat{\Lambda}^{12} - C_{312} \partial_0 \hat{\Lambda}^{12} + C_{[03]} \partial_1 \partial_2 \hat{\Lambda}^{12} \right. \\
    &\qquad \quad + \left( \partial_0 \Lambda_{12} - \partial_1 \partial_2 \Lambda_0 \right) \hat{C}^{12}_3 - \left( \partial_3 \Lambda_{12} - \partial_1 \partial_2 \Lambda_3 \right) \hat{C}^{12}_0 + \left( \partial_0 \Lambda_3 - \partial_3 \Lambda_0 \right) \hat{C} \\
    &\qquad  \left. + \left( \partial_0 \Lambda_{12} - \partial_1 \partial_2 \Lambda_0 \right) \partial_3 \hat{\Lambda}^{12} - \left( \partial_3 \Lambda_{12} - \partial_1 \partial_2 \Lambda_3 \right) \partial_0 \hat{\Lambda}^{12} + \left( \partial_0 \Lambda_3 - \partial_3 \Lambda_0 \right) \partial_1 \partial_2 \hat{\Lambda}^{12} \right)  \\
    &= \frac{iN}{2\pi} \partial_3 \left[ \hat{\Lambda}^{12} C_{012} - \Lambda_0 \hat{C} -\Lambda_{12} \hat{C}^{12}_0 - \Lambda_0 \partial_1 \partial_2 \hat{\Lambda}^{12} - \Lambda_{12} \partial_0 \hat{\Lambda}^{12}  \right] \,.
\end{split}
\end{align}
Thus on the boundary $x^3 = 0$, the term 
\begin{align}
    \delta_g S_{\text{SSPT,e}} = -\int dx^0 dx^1 dx^2 \frac{iN}{2\pi}  \left[ \hat{\Lambda}^{12} C_{012}  -\Lambda_{12} ( \hat{C}^{12}_0 + \partial_0 \hat{\Lambda}^{12} ) - \Lambda_0 ( \hat{C} + \partial_1 \partial_2 \hat{\Lambda}^{12} )   \right]_{x^3 =0} \label{eboundarysspt}
\end{align}
arises. This boundary term matches the 't Hooft anomaly of the 2+1d exotic $BF$ theory \eqref{eanomaly} on the boundary $x^3=0$. Therefore we can cancel the 't Hooft anomaly of the 2+1d exotic $BF$ theory on the boundary by the gauge-variation of the 3+1d SSPT phase on the bulk.\footnote{The anomaly \eqref{eanomaly} and the boundary term \eqref{eboundarysspt} have the same sign, so we have to consider the SSPT partition function $Z_{\text{SSPT,e}} =\int d\beta d\hat{\beta} \,e^{S_{\text{SSPT,e}}}$.}

\section{Anomaly in the 2+1d Foliated $BF$ Theory} \label{section 3}

In this section, we first review the foliated $BF$ theory in 2+1 dimensions \cite{Ohmori:2022rzz} and then consider its mixed 't Hooft anomaly. The mixed 't Hooft anomaly is considered to be the same one as the exotic $BF$ theory in 2+1 dimensions in Section \ref{section 2} from the foliated-exotic duality. To see this anomaly, we have to couple the subsystem symmetries of the foliated $BF$ theory to background gauge fields, but in the foliated form, its construction is non-trivial. In Section \ref{section 32}, we will determine field correspondences between background tensor gauge fields and background foliated gauge fields, and construct the foliated $BF$ Lagrangian coupled to the foliated background gauge fields. Along the way, we will discuss a new type of symmetry-like operator. Next, in Section \ref{section 33}, we will construct the foliated SSPT phase in 3+1 dimensions from the exotic SSPT phase in Section \ref{section 23} using the field correspondences. The foliated SSPT phase is the foliated form of the exotic SSPT phase, so then we will have established the foliated-exotic duality of the 3+1d SSPT phase.

\subsection{Foliated $BF$ Theory and Field Correspondences}
\label{section 31}

Again we take a three-torus of lengths $l^0$, $l^1$, $l^2$ as a Euclidean spacetime and the coordinates $(x^0,x^1,x^2)$ on it. We consider a $BF$ theory on the two-dimensional spatial manifold that is regarded as a stack of an infinite number of one-dimensional spatial submanifolds. These submanifolds are called leaves and such a structure of a decomposition of a manifold is called a codimension-one foliation. A QFT on such a manifold is called a foliated QFT (FQFT) \cite{Slagle:2018swq,Slagle:2020ugk,Hsin:2021mjn}. A codimension-one foliation is characterized by a one-form foliation field $e$, which is orthogonal to the leaves. Here we consider two simultaneous flat foliations $e^k = dx^k \, (k =1,2)$, where the indices $k$ indicate the direction of the foliations.

The foliated $BF$ theory in 2+1 dimensions contains two types of foliated gauge fields, which are regarded as gauge fields on the leaves, and bulk gauge fields mediating the foliated gauge fields on each leaf \cite{Slagle:2018swq,Slagle:2020ugk,Hsin:2021mjn,Ohmori:2022rzz}.\footnote{The word “bulk” in the bulk gauge field has nothing to do with the word “bulk” in the bulk SSPT phase.} The foliated gauge fields are $U(1)$ foliated $A$-type (1+1)-form gauge fields $A^k \wedge dx^k$ and $U(1)$ foliated $B$-type zero-form gauge fields $B^k$.\footnote{We use the words \textit{A-type} and \textit{B-type} used in \cite{Ohmori:2022rzz}.} The bulk gauge fields are $U(1)$ one-form gauge fields $a$ and $b$. The gauge transformations of the foliated gauge fields are
\begin{align}
    A^k \wedge dx^k &\rightarrow A^k \wedge dx^k +d \zeta^k \wedge dx^k  \, , \\
    \hat{B}^k &\rightarrow \hat{B}^k +2\pi \hat{t}^k - \mu \, ,
\end{align}
where $\zeta^k \wedge dx^k$ is a (0+1)-form gauge parameter, $\hat{t}^k$ is an $x^k$-dependent integer-valued gauge parameter, and $\mu$ is a zero-form bulk gauge parameter. 
The gauge parameter $\zeta^k \wedge dx^k$ has its own gauge transformation $\zeta^k \wedge dx^k \rightarrow \zeta^k \wedge dx^k + 2\pi d \xi^k$, where $\xi^k$ is an $x^k$-dependent integer-valued gauge parameter. The gauge transformations of the bulk gauge fields are
\begin{subequations}
 \begin{align}
     a &\rightarrow a + d\kappa - \sum^{2}_{k=1} \zeta^k dx^k\,, \label{fgauge3} \\
     b &\rightarrow b + d\mu\,, \label{fgauge4}
 \end{align}
 \end{subequations}
where $\kappa$ are zero-form bulk gauge parameters that have their own gauge transformations. The gauge transformation of $\kappa$ is $\kappa \rightarrow \kappa + 2\pi \xi^1 + 2\pi \xi^2$, where $\xi^k$ are the parameters for the transformation of $\zeta^k$. Note that the constant modes of $\xi^k$ make $\kappa$ a $U(1)$-valued parameter. These foliated gauge fields and bulk gauge fields can have particular types of singularities and discontinuities. (See \cite{Ohmori:2022rzz} for more details.)

The foliated BF Lagrangian is
\begin{align}
    \L_{\text{f}} = \frac{iN}{2\pi} \sum^{2}_{k=1}  (d\hat{B}^k + b)\wedge A^k \wedge dx^k + \frac{iN}{2\pi} b \wedge da\,, \label{21folilag}
\end{align}
where $N$ is an integer. Due to the $BF$ couplings, these $U(1)$ gauge fields are Higgsed down to $\Z_N$.

The equations of motion are
\begin{align}
     \frac{N}{2\pi}(d\hat{B}^k + b) \wedge dx^k = 0\,, \label{feom1}\\
     \frac{N}{2\pi}db = 0\,, \label{feom2}\\
     \frac{N}{2\pi}dA^k \wedge dx^k = 0\,, \label{feom3}\\
     \frac{N}{2\pi} \left( \sum^{2}_{k=1}  A^k \wedge dx^k +  da \right)  =0\,. \label{feom4}
\end{align}

This foliated $BF$ theory is equivalent to the exotic $BF$ theory under field correspondences, which is called a foliated-exotic duality \cite{Ohmori:2022rzz,Spieler:2023wkz}. To show this duality, we must integrate out the time component of the fields $b_0$, $A^1_0$ and $A^2_0$ in the foliated $BF$ theory, then solve the equations of motion for $b_1$ and $b_2$, and plug them into the Lagrangian. However, from the form of the couplings, this manipulation leads to the same result as integrating out $b_1$ and $b_2$ instead of $A^1_0$ and $A^2_0$, then solving the equations for $A^1_0$ and $A^2_0$, and plugging them in. Here we integrate out $b$ for simplicity, and then we can use the equation of motion \eqref{feom4}, or in components,
\begin{align}
    \frac{N}{2\pi} (A^1_0 + \partial_0 a_1 - \partial_1 a_0) &= 0 \,, \label{aeomcomp1}\\
    \frac{N}{2\pi} (A^2_0 + \partial_0 a_2 - \partial_2 a_0) &= 0 \,, \label{aeomcomp2} \\
    \frac{N}{2\pi} (A^1_2 - A^2_1 + \partial_2 a_1 - \partial_1 a_2) &= 0 \,. \label{aeomcomp3} 
\end{align}
The correspondences between the tensor gauge fields and the foliated and bulk gauge fields are\footnote{The symbol $\simeq$ means that the correspondence between the gauge fields or parameters in the exotic theory and the foliated theory.}
\begin{align}
    A_0 &\simeq a_0 \,,  \label{acorr}\\
    \partial_k A_0 &\simeq A^k_0 + \partial_0 a_k \,,\quad (k=1,2) \,, \label{ak0corr} \\
    A_{12}  &\simeq A^1_2 + \partial_2 a_1 = A^2_1 + \partial_1 a_2 \,, \label{a12corr} \\
    \hat{\phi}^{12} &\simeq \hat{B}^1 - \hat{B}^2  \,. \label{bcorr}
\end{align}
Note that the gauge transformations $\zeta^k dx^k$ and $\mu$ in the right hand sides cancel out, so the degrees of freedom of the fields are consistent. In addition, we have correspondences between the gauge parameters
\begin{gather}
    \alpha \simeq \kappa \,, \\
    \hat{m}^k  \simeq \hat{t}^k \,, \\
    n^k \simeq \xi^k  \,. 
\end{gather}
Then, the exotic $BF$ Lagrangian can be transformed to the foliated $BF$ Lagrangian after integrating out $b$:
\begin{align}
\begin{split}
    \L_{\text{e}} &= \frac{i N}{2\pi}\hat{\phi}^{12}(\partial_0 A_{12} - \partial_{1}\partial_{2} A_0) \\
    &\simeq \frac{i N}{2\pi} (\hat{B}^1 - \hat{B}^2) (\partial_0 A_{12} - \partial_{1}\partial_{2} A_0) \\
    &\simeq \frac{i N}{2\pi} \hat{B}^1 \left[ \partial_0 (A^1_2 + \partial_2 a_1 )  - \partial_{2} ( A^1_0 + \partial_0 a_1 ) \right] \\ & \qquad - \frac{i N}{2\pi} \hat{B}^2 \left[ \partial_0 (A^2_1 + \partial_1 a_2 )  - \partial_{1} ( A^2_0 + \partial_0 a_2 ) \right] \\
    &= \frac{iN}{2\pi} \left[ -\partial_0 \hat{B}^1  A^1_2 + \partial_2 \hat{B}^1 A^1_0 + \partial_0 \hat{B}^2 A^2_1 - \partial_1 \hat{B}^2 A^2_0 \right]  \\
    &= \frac{iN}{2\pi} \sum^{2}_{k=1}  d\hat{B}^k \wedge A^k \wedge dx^k\,.
\end{split}
\end{align}
We return $b$, and then we get the foliated $BF$ Lagrangian
\begin{align}
    \L_{\text{f}} = \frac{iN}{2\pi} \sum^{2}_{k=1}  d\hat{B}^k \wedge A^k \wedge dx^k + \frac{iN}{2\pi} b \wedge  \left( \sum^{2}_{k=1} A^k \wedge dx^k + da \right)\,.
\end{align}

Using the field correspondences, we can also derive the gauge-invariant operators and defects in the foliated $BF$ theory. 
The symmetry operator associated with the $\Z_N$ electric global symmetry is
\ali{
 V[x] = \exp \left[ i (\hat{B}^1 - \hat{B}^2 ) \right] \,. \label{felesym}
}
The strip operators associated with the $\Z_N$ dipole global symmetries are
\ali{
    W_k[S^k_2] = \exp \left[ i \oint_{S^k_2} \left( A^k  \wedge dx^k + d(a_k dx^k) \right) \right] \,, \quad (k=1,2) \,. \label{modstrip}
}
The quadrupole operator associated with the $\Z_N$ tensor time-like symmetry is
\ali{
T\left[C_1^{12,\text{rect}}(x^1_1,x^1_2,x^2_1,x^2_2)\right] = \exp\left[-i \Delta_{12} ( \hat{B}^1 - \hat{B}^2) (x^1_1,x^1_2,x^2_1,x^2_2) \right] \,. \label{ftimesym}
}
The fracton defect is
\ali{
F[C_1^0] = \exp \left[ i \oint_{C_1^0} a \right]\,. \label{ffracton}
}

\subsection{Coupling to the Background Foliated Gauge Fields}
\label{section 32}

In Section \ref{section 22}, we coupled the symmetries in the exotic $BF$ theory to the background tensor gauge fields. Since the foliated $BF$ theory is equivalent to the exotic $BF$ theory, it should be possible to couple the foliated $BF$ theory to the same set of the background gauge fields. However, in the case of the foliated theory, the structure of symmetry operators seems to be more complicated and it is non-trivial to find appropriate couplings. Here we construct the foliated $BF$ Lagrangian including background foliated and bulk gauge fields using the field correspondences in the foliated-exotic duality.

The exotic $BF$ Lagrangian coupled to background tensor gauge fields is
\sbali{
    \L_{\text{e}} \left[ C_{012}, \hat{C}_0^{12}, \hat{C} \right] &=  \L_{\text{e},BF}  + \L_{\text{e}, \hat{\chi}} + \L_{\text{e},\chi} \,,\\
    \L_{\text{e},BF} &= \frac{iN}{2\pi} \left[ \hat{\phi}^{12} (\partial_0 A_{12} - \partial_1 \partial_2 A_0 -C_{012} ) + A_{12} \hat{C}^{12}_0 + A_0 \hat{C} \right] \,, \\ 
    \L_{\text{e},\hat{\chi}} &= \frac{iN}{2\pi}\hat{\chi}^{12} C_{012}\,, \\
    \L_{\text{e},\chi} &= \frac{iN}{2\pi} \chi (\partial_0 \hat{C} - \partial_1 \partial_2 \hat{C}^{12}_0)  \,.
}
Firstly, we consider the $BF$ part $\L_{\text{e},BF}$. Under the equations of motion \eqref{aeomcomp1}--\eqref{aeomcomp3}, the exotic Lagrangian is equivalent to the foliated Lagrangian. When coupled to background gauge fields, we assume that the background gauge transformations of the foliated gauge fields are
\begin{align}
 A^k\wedge dx^k &\rightarrow A^k \wedge dx^k + \lambda^k \wedge dx^k  \,,  \label{akbk} \\
 a &\rightarrow a + \lambda \,, \label{abk} \\
 \hat{B}^k &\rightarrow \hat{B}^k + \hat{\lambda}^k  \,,\label{bkbk} 
\end{align}
where $\lambda^k \wedge dx^k$ is a (1+1)-form gauge parameter, $\lambda$ is a one-form gauge parameter and $\hat{\lambda}^k$ is a zero-form gauge parameter. For the background gauge invariance, we demand that the $b$ equations of motion be
\begin{align}
    \frac{iN}{2\pi} (A^1_0 + \partial_0 a_1 - \partial_1 a_0 - c_{01}) &= 0 \,, \\
    \frac{iN}{2\pi} (A^2_0 + \partial_0 a_2 - \partial_2 a_0  - c_{02}) &= 0 \,, \\
    \frac{iN}{2\pi}(A^1_2 - A^2_1 + \partial_2 a_1 - \partial_1 a_2  + c_{12})  &= 0 \,,
\end{align}
instead of \eqref{aeomcomp1}--\eqref{aeomcomp3}, or in the differential form,
\ali{
    \frac{iN}{2\pi}\left( \sum_{k=1}^2  A^k \wedge dx^k + da - c \right) = 0 \,, 
}
where $c$ is the two-form gauge field that has a background gauge transformation
\ali{
    c \rightarrow c + d\lambda + \sum_{k=1}^2 \lambda^k \wedge dx^k \,. \label{cbk}
}
Then, we have the new field correspondences including $c$
\ali{
    A_0 &\simeq a_0 \,, \\
    \partial_k A_0 &\simeq A^k_0 + \partial_0 a_k - c_{0k} \,, \quad (k=1,2)\,, \\
    A_{12}  &\simeq A^1_2 + \partial_2 a_1 = A^2_1 + \partial_1 a_2  - c_{12}  \,, \\
    \hat{\phi}^{12} &\simeq \hat{B}^1 - \hat{B}^2  \,.
}
Due to $c_{12}$, the $k=1$ and $k=2$ foliated gauge fields are not treated symmetrically. From \eqref{a0bk}, \eqref{a12bk}, \eqref{phi12bk}, \eqref{akbk}--\eqref{bkbk} and \eqref{cbk}, we derive correspondences between the background gauge parameters
\begin{align}
    \Lambda_0 &\simeq \lambda_0 \,, \\
    \Lambda_{12} &\simeq \lambda^1_2 + \partial_2 \lambda_1 \,, \\
    \hat{\Lambda}^{12} &\simeq \hat{\lambda}^1 - \hat{\lambda}^2 \,.
\end{align}
Note that $\lambda^2_1$ and $\lambda_2$ do not appear.

Using the field correspondences, $\L_{\text{e},BF}$ can be written as
\ali{
\L_{\text{e},BF} \simeq  \frac{iN}{2\pi} \left[ (\hat{B}^1 - \hat{B}^2) (\partial_0 A^1_2 -  \partial_2 A^1_0 + \partial_2 c_{01} - C_{012} ) + (A^1_2 + \partial_2 a_1) \hat{C}^{12}_0 + a_0 \hat{C} \right] \,,
}
where we can replace $\partial_0 A^1_2 -  \partial_2 A^1_0 + \partial_2 c_{01}$ and $A^1_2 + \partial_2 a_1$ with $\partial_0 A^2_1 -  \partial_1 A^2_0 + \partial_1 c_{02} - \partial_0 c_{12}$ and $A^2_1 + \partial_1 a_2 - c_{12} $ respectively.

Since $A^k \wedge dx^k$, $\hat{B}^k$ and $a$ are the foliated gauge fields and the bulk gauge fields in the foliated theory, we want to substitute some background foliated and bulk gauge fields for the background tensor gauge fields $C_{012}$ and $(\hat{C}^{12}_0, \hat{C})$. Therefore we introduce a $U(1)$ foliated $A$-type (2+1)-form background gauge field $C^k \wedge dx^k\, (k=1,2)$, a $U(1)$ foliated $B$-type one-form background gauge field $\hat{C}^k \, (k=1,2)$ that obeys $\hat{C}^k_k = 0$, and a bulk two-form background gauge field $\hat{c}_{12} \, dx^1 dx^2$. Then, we assume correspondences between the background gauge fields including $c$ as\footnote{Under the $90$ degree rotation $x^1 \rightarrow x^2$, $x^2 \rightarrow -x^1$, the $C$ fields transform as $C_{012} \rightarrow -C_{012}$ and $C^1_{02} \leftrightarrow - C^2_{01}$, and the $\hat{C}$ fields transform as $\hat{C}^{12}_0 \rightarrow -\hat{C}^{12}_0$, $\hat{C} \rightarrow \hat{C}$, $\hat{C}^1_0 \leftrightarrow \hat{C}^2_0$, $\hat{C}^1_2 \rightarrow -\hat{C}^2_1$ and $\hat{C}^2_1 \rightarrow \hat{C}^1_2$. Thus the rotational transformations on the both sides are compatible.}
\ali{
C_{012} &\simeq C^1_{02} + \partial_2 c_{01} = C^2_{01} + \partial_1 c_{02} - \partial_0 c_{12}  \,, \label{ccorr1}  \\
\hat{C}_0^{12} &\simeq \hat{C}^1_0 - \hat{C}^2_0 \,,  \label{chatcorr1} \\
\hat{C} &\simeq \partial_1 \hat{C}^1_2 - \partial_2 \hat{C}^2_1 + \hat{c}_{12} \,. \label{chatcorr2}
}
Note that the field correspondence of $\hat{C}$ is the same form as that of $\hat{B}$ in the foliated $BF$ theory with two foliations in 3+1 dimensions\cite{Spieler:2023wkz}. To impose $C^1_{02} + \partial_2 c_{01}  = C^2_{01} + \partial_1 c_{02} - \partial_0 c_{12}$, we must add to the Lagrangian the term $\frac{iN}{2\pi} h  \sum_{k=1}^2 ( C^k \wedge dx^k - dc) $ where $h$ is a zero-form dynamical field, which has a dynamical gauge transformation
\ali{
h \rightarrow h + 2\pi w + \mu\,, 
}
where $w$ is an integer-valued gauge parameter, which is canceled by the gauge transformation of $\hat{\chi}^k$ mentioned later. The background gauge transformations of $C^k \wedge dx^k$, $\hat{C}^k$ and $\hat{c}_{12}$ are
\ali{
    C^k \wedge dx^k &\rightarrow C^k \wedge dx^k + d\lambda^k \wedge dx^k    \,, \\
    \hat{C}^k &\rightarrow \hat{C}^k + d\hat{\lambda}^k - \hat{\nu}   \,, \\
    \hat{c}_{12} &\rightarrow \hat{c}_{12} + (d\hat{\nu})_{12} \,, 
}
where $\hat{\nu}$ is a one-form gauge parameter.
Under the assumption, $\L_{\text{e},BF}$ can be written as
\alis{
\L_{\text{e},BF} \simeq & \frac{iN}{2\pi} \left[ \hat{B}^1 (\partial_0 A^1_2 -  \partial_2 A^1_0 -C^1_{02} ) - \hat{B}^2 (\partial_0 A^2_1 -  \partial_1 A^2_0 -C^2_{01} )\right. \\
& \qquad \left. + (A^1_2 + \partial_2 a_1) \hat{C}^1_0 - (A^2_1 + \partial_1 a_2 - c_{12})\hat{C}^2_0 + a_0 (\partial_1 \hat{C}^1_2 - \partial_2 \hat{C}^2_1 + \hat{c}_{12})\right]  \,.
}
Integrating the $\hat{C}^k_i$ term by parts, we substitute 
\alis{
a_0 (\partial_1 \hat{C}^1_2 - \partial_2 \hat{C}^2_1 ) &\rightarrow -\partial_1 a_0 \, \hat{C}^1_2  + \partial_2 a_0\, \hat{C}^2_1 \\
&\simeq -(A^1_0 + \partial_0 a_1 - c_{01} )\hat{C}^1_2 + (A^2_0 + \partial_0 a_2 - c_{02} ) \hat{C}^2_1 \,.
}
Then, we can obtain 
\alis{
\L_{\text{e},BF} \simeq & \frac{iN}{2\pi} \sum_{k=1}^2\left[ - \hat{B}^k\,(dA^k - C^k) \wedge dx^k - \hat{C}^k \wedge \left(A^k \wedge dx^k + d(a_k dx^k) \right) \right] \\
 &+ \frac{iN}{2\pi} a_0 \hat{c}_{12}\, d^3 x +\frac{iN}{2\pi} ( c_{01} \hat{C}^1_2 - c_{02} \hat{C}^2_1 - c_{12} \hat{C}^2_0 ) d^3 x \,,
}
where $d^3 x = dx^0 dx^1 dx^2$. We will see later that the term 
\ali{
\frac{iN}{2\pi} \left( c_{01} \hat{C}^1_2 - c_{02} \hat{C}^2_1 - c_{12} \hat{C}^2_0 \right) d^3 x \label{dropterm}
}
can be dropped by combining the bulk SSPT phase as a counterterm, so we drop it here. Restoring the $b$ term and adding the $h$ term, we get the foliated $BF$ Lagrangian including the background gauge fields:
\alis{
\L_{\text{f},BF} = &\frac{iN}{2\pi} \sum_{k=1}^2\left[ - \hat{B}^k\,(dA^k - C^k) \wedge dx^k - \hat{C}^k \wedge \left(A^k \wedge dx^k + d(a_k dx^k) \right)   \right]  \\
& + \frac{iN}{2\pi} a_0 \hat{c}_{12}\, d^3 x  + \frac{iN}{2\pi} b \wedge \left( \sum_{k=1}^2 A^k \wedge dx^k + da - c \right)+ \frac{iN}{2\pi} h  \left( \sum_{k=1}^2 C^k \wedge dx^k - dc \right) \,,
}
or integrating it by parts, 
\alis{
\L_{\text{f},BF} = &\frac{iN}{2\pi} \sum_{k=1}^2\left[  (d \hat{B}^k - \hat{C}^k + b)  \wedge A^k \wedge dx^k + \hat{B}^k\,C^k \wedge dx^k - \hat{C}^k \wedge  d(a_k dx^k)   \right] \\
& + \frac{iN}{2\pi} a_0 \hat{c}_{12}\, d^3 x + \frac{iN}{2\pi} b \wedge ( da - c ) + \frac{iN}{2\pi} h \left( \sum_{k=1}^2 C^k \wedge dx^k - dc \right)  \,.
}
Then, $b$ has the background gauge transformation
\ali{
b \rightarrow b - \hat{\nu} \,, \label{bpass}
}
which is canceled by the background gauge transformation $\hat{\nu}$ of $\hat{C}^k$ and $\hat{c}$ up to a background term as
\alis{
   \delta_{\hat{\nu}} &\left\{ \sum_{k=1}^2 \left[ (-\hat{C}^k + b ) \wedge A^k \wedge dx^k - \hat{C}^k \wedge  d(a_k dx^k) \right] + a_0 \hat{c}_{12}\, d^3 x  + b \wedge (da - c) \right\} \\
   &= \sum_{k=1}^2   \hat{\nu} \wedge  d(a_k dx^k) + a_0 (d \hat{\nu})_{12}\, d^3 x  - \hat{\nu} \wedge ( da - c)  \\
   &= \hat{\nu} \wedge  c  \,.
}
The background foliated gauge fields $C^k \wedge dx^k$ are coupled to the $\Z_N$ electric symmetry \eqref{felesym}, the background bulk gauge field $c$ is coupled to the $\Z_N$ tensor time-like symmetry \eqref{ftimesym}, and the background foliated gauge fields $\hat{C}^k$ are coupled to the $\Z_N$ dipole symmetries \eqref{modstrip}.

From the term $\frac{iN}{2\pi} a_0 \hat{c}_{12}\, d^3 x$, we can say the background gauge fields $ \hat{c}_{12} \, dx^1 dx^2 $ is coupled to the fracton defect $F[C_1^0] = \exp \left[ i \oint_{C_1^0} a \right]$. In addition, the quadrupole operator associated with the $\Z_N$ tensor time-like symmetry \eqref{ftimesym} can be written as
\alis{
    T\left[C_1^{12,\text{rect}}(x^1_1,x^1_2,x^2_1,x^2_2)\right] &= \exp\left[-i \Delta_{12} ( \hat{B}^1 - \hat{B}^2) (x^1_1,x^1_2,x^2_1,x^2_2) \right] \\
    &= \exp\left[i \oint_{C_1^{12,\text{rect}}} b \right] \,
}
by using the equations of motion \eqref{feom1}, and then the symmetry action \eqref{tfctvw} can be written as
\ali{
    T\left[C_1^{12,\text{rect}}(x^1_1,x^1_2,x^2_1,x^2_2)\right] \cdot F[C_1^0] = e^{-2\pi i/N} F[C_1^0]
}
in the foliated form \cite{Slagle:2020ugk,Ohmori:2022rzz}. Actually, $T\left[C_1^{12,\text{rect}}(x^1_1,x^1_2,x^2_1,x^2_2)\right]$ acts on the gauge field $b$, but we can formally interpret the background gauge transformation of $b$ \eqref{bpass} as being passively acted by the defect $F[C_1^0]$. If $\hat{\nu}$ were not a local transformation, this transformation would not be a symmetry transformation in the original 2+1d foliated $BF$ theory \eqref{21folilag}. Moreover, the fracton defect $F[C_1^0]$ is not topological, so the defect is not a symmetry operator. However, this situation is similar to that of global symmetries, so we call the fracton defect a \textit{symmetry-like passive action operator}.

The remaining $\hat{\chi}^{12}$ and $\chi$ terms can also be written as foliated forms. As for $\L_{\text{e},\hat{\chi}}$, we assume correspondences between the dynamical field $\hat{\chi}^{12}$ as 
\ali{
&\hat{\chi}^{12} \simeq \hat{\chi}^1 - \hat{\chi}^2 \,,
}
where $\hat{\chi}^k\, (k=1,2)$ are $x^k$-dependent integer-valued dynamical fields. Their dynamical gauge transformations are
\ali{
    \hat{\chi}^k &\rightarrow \hat{\chi}^k + 2\pi \hat{t}^k + 2\pi w \,,
}
where $\hat{t}^k$ and $w$ are canceled by the gauge transformations of $\hat{B}^k$ and $h$ respectively. 
Then, we rewrite $\L_{\text{e},\hat{\chi}}$ as
\alis{
 \L_{\text{e},\hat{\chi}} &= \frac{iN}{2\pi}\hat{\chi}^{12} C_{012} \\ 
 &\simeq \frac{iN}{2\pi} \left[ \hat{\chi}^1 ( C^1_{02} + \partial_2 c_{01} ) - \hat{\chi}^2 (C^2_{01} + \partial_1 c_{02} - \partial_0 c_{12}  ) \right] \\
 &= -\frac{iN}{2\pi} \sum_{k=1}^2  \hat{\chi}^k C^k \wedge dx^k  \,.
}
As for $\L_{\text{e},\chi}$, we use $\chi$ and introduce dynamical (0+1)-form fields $\chi^k dx^k\, (k=1,2)$ as
\ali{
    \partial_k \chi = \chi^k \,. \label{chiconst}
}
Their dynamical gauge transformations are
\ali{
    \chi &\rightarrow \chi + \kappa \,, \\
    \chi^k &\rightarrow \chi^k + \partial_k \kappa \,.
}
Then, we can rewrite $\L_{\text{e},\chi}$ as
\alis{
\L_{\text{e},\chi} &= \frac{iN}{2\pi} \chi (\partial_0 \hat{C} - \partial_1 \partial_2 \hat{C}^{12}_0) \\
&\simeq \frac{iN}{2\pi} \chi \left[ \partial_1 (\partial_0 \hat{C}^1_2 - \partial_2 \hat{C}^1_0) - \partial_2 (\partial_0 \hat{C}^2_1 - \partial_1 \hat{C}^2_0) + \partial_0 \hat{c}_{12} \right] \\
&= \frac{iN}{2\pi}  \left[ \partial_1 \chi(-\partial_0 \hat{C}^1_2 + \partial_2 \hat{C}^1_0) + \partial_2 \chi (\partial_0 \hat{C}^2_1 - \partial_1 \hat{C}^2_0) + \chi \partial_0 \hat{c}_{12} \right] \\
&= \frac{iN}{2\pi} \left( \sum_{k=1}^2  \chi^k  dx^k \wedge  d\hat{C}^k  + \chi \partial_0 \hat{c}_{12}\, d^3x  \right)  \,.
}
To impose \eqref{chiconst}, we should add a term of dynamical bulk gauge fields $\hat{c}_{01} dx^0 dx^1$ and $\hat{c}_{02} dx^0 dx^2$, and then we have the form
\alis{
\L_{\text{f},\chi} 
&= \frac{iN}{2\pi} \left( \sum_{k=1}^2  \chi^k  dx^k \wedge  d\hat{C}^k  + \chi \partial_0 \hat{c}_{12}\, d^3x  \right)  \\ & \quad + \frac{iN}{2\pi} \left[ \hat{c}_{01} ( \chi^2 - \partial_2 \chi ) - \hat{c}_{02} ( \chi^1 - \partial_1 \chi ) \right] d^3x  \\
&= \frac{iN}{2\pi}\sum_{k=1}^2  \left[  \chi^k  dx^k \wedge  ( d\hat{C}^k + \hat{c} ) + \chi\, d \hat{c}  \right] \,,
}
where the background gauge transformations of $\hat{c}_{01}$ and $\hat{c}_{02}$ are
\ali{
   \hat{c}_{01} &\rightarrow \hat{c}_{01} + (d\hat{\nu})_{01} \,, \\
   \hat{c}_{02} &\rightarrow \hat{c}_{02} + (d\hat{\nu})_{02} \,.
}

Note that we combine the background gauge field $\hat{c}_{12}$ and the dynamical gauge fields $\hat{c}_{01}$ and $\hat{c}_{02}$ into $\hat{c}$. To obtain the exotic theory, we must integrate out $\hat{c}_{01}$ and $\hat{c}_{02}$.

After all, we have constructed the 2+1d foliated $BF$ Lagrangian coupled to the background gauge fields. The full Lagrangian is
\alis{
&\L_{\text{f}} \left[ C^k \wedge dx^k , \hat{C}^k, c, \hat{c}_{12} \right] \\
& \qquad = \frac{iN}{2\pi} \sum_{k=1}^2\left[  (d \hat{B}^k - \hat{C}^k + b)  \wedge A^k \wedge dx^k + \hat{B}^k\,C^k \wedge dx^k - \hat{C}^k \wedge  d(a_k dx^k)  \right] \\
&\qquad  + \frac{iN}{2\pi} a_0 \hat{c}_{12}\, d^3 x  + \frac{iN}{2\pi} b \wedge ( da - c ) + \frac{iN}{2\pi} h  \left( \sum_{k=1}^2 C^k \wedge dx^k - dc \right)  \\
& \qquad  + \frac{iN}{2\pi} \sum_{k=1}^2 \left[ - \hat{\chi}^k C^k \wedge dx^k  +  \chi^k  dx^k \wedge  ( d\hat{C}^k + \hat{c} ) + \chi\, d \hat{c}  \right]  \,. \label{fbfc}
}
Under the background gauge transformations, it transforms as
\alis{
\delta_g \L_{\text{f}} 
= &\frac{iN}{2\pi} \sum_{k=1}^2\left[  (d \hat{B}^k - \hat{C}^k + b)  \wedge \lambda^k \wedge dx^k + \hat{B}^k d\lambda^k \wedge dx^k +  \hat{\lambda}^k ( C^k  + d\lambda^k )\wedge dx^k \right. \\
& \left. - \hat{C}^k  \wedge  d( \lambda_k dx^k) - (d\hat{\lambda}^k - \hat{\nu}) \wedge  d(a_k dx^k + \lambda_k dx^k)  \right] \\
& + \frac{iN}{2\pi} a_0 (d\hat{\nu})_{12}\, d^3 x  + \frac{iN}{2\pi} \lambda_0 (\hat{c} + d\hat{\nu})_{12}  \, d^3 x  \\
& + \frac{iN}{2\pi}b \wedge \left(  - \sum^2_{k=1} \lambda^k \wedge dx^k  \right) - \frac{iN}{2\pi} \hat{\nu} \wedge \left( da - c - \sum^2_{k=1} \lambda^k \wedge dx^k \right) \\
= & \frac{iN}{2\pi} \sum_{k=1}^2\left[ \hat{\lambda}^k \, C^k  \wedge dx^k  - ( \hat{C}^k + d \hat{\lambda}^k)  \wedge \left\{\lambda^k \wedge dx^k + d(\lambda_k dx^k) \right\} \right] + \frac{iN}{2\pi} \lambda_0 \hat{c}_{12}\, d^3 x  \\
& + \frac{iN}{2\pi} \hat{\nu} \wedge \left( c + d\lambda + \sum^2_{k=1} \lambda^k \wedge dx^k \right) \,. \label{fanomaly}
}
Note that to derive this formula from the exotic one \eqref{eanomaly} by using the field correspondences, we have to take into account the dropped term $\frac{iN}{2\pi} ( c_{01} \hat{C}^1_2 - c_{02} \hat{C}^2_1 - c_{12} \hat{C}^2_0 )$.

\subsection{Foliated SSPT Phase in 3+1 Dimensions}
\label{section 33}

In Section \ref{section 23}, we saw that the mixed 't Hooft anomaly in the 2+1d exotic $BF$ theory are canceled by the 3+1d exotic SSPT phase. In this section, we construct a description of the foliated SSPT phase that is equivalent to the exotic one with two foliations by determining field correspondences. Again we take the coordinates $(x^0,x^1,x^2,x^3)$.

First, we introduce background foliated gauge fields $C^k \wedge dx^k\, (k = 1,2)$ and a background bulk gauge field $c$, and assume correspondences between the background tensor gauge fields $(C_{012}, C_{312}, C_{[03]})$ in the exotic SSPT phase and $C^k \wedge dx^k$, $c$ as
\ali{
    C_{012} &\simeq C^1_{02} + \partial_2 c_{01} = C^2_{01} + \partial_1 c_{02} - \partial_0 c_{12} \,, \label{3cor1} \\
    C_{312} &\simeq C^1_{32} + \partial_2 c_{31} = C^2_{31} + \partial_1 c_{32} - \partial_3 c_{12} \label{3cor2} \,, \\
    C_{[03]} &\simeq c_{03} \label{3cor3} \,, \\
    \partial_1 C_{[03]} &\simeq C^1_{03} + \partial_3 c_{01}  -  \partial_0 c_{31} \,, \label{3cor4}\\
    \partial_2 C_{[03]} &\simeq C^2_{03} + \partial_3 c_{02} - \partial_0 c_{32} \label{3cor5}   \,, 
}
where $C^k \wedge dx^k$ is a $U(1)$ foliated $A$-type (2+1)-form gauge field and $c$ is a $U(1)$ bulk two-form gauge field. These correspondences are consistent with the ones between the background gauge fields in the 2+1d exotic and foliated $BF$ theories. To impose the constraints for $C^k \wedge dx^k$ and $c$, we add the term $\frac{iN}{2\pi} p \wedge  \left(\sum_{k=1}^2 C^k \wedge dx^k - dc \right)$ to the Lagrangian where $p$ is a dynamical one-form field. $p$ has a dynamical gauge transformation $p \rightarrow p + d\hat{v}$, where $\hat{v}$ is a zero-form gauge parameter. The background gauge transformations of $C^k \wedge dx^k$ and $c$ are
\ali{
    C^k \wedge dx^k &\rightarrow C^k \wedge dx^k + d\lambda^k \wedge dx^k \,, \quad (k=1,2) \,, \\
    c &\rightarrow c + d\lambda + \sum_{k=1}^2 \lambda^k \wedge dx^k \,,
}
where $\lambda^k \wedge dx^k$ is a (1+1)-form gauge parameter and $\lambda$ is a one-form gauge parameter. We have correspondences between background gauge parameters:
\ali{
    \Lambda_0 &\simeq \lambda_0 \,, \\
    \Lambda_3 &\simeq \lambda_3 \,, \\
    \Lambda_{12} &\simeq \lambda^1_2 + \partial_2 \lambda_1  \,.
}

Next, we introduce background foliated gauge fields $\hat{C}^k\, (k = 1,2)$ and a background bulk gauge field $\hat{c}$, and assume correspondences between the background tensor gauge fields $(\hat{C}^{12}_{0}, \hat{C}, \hat{C}^{12}_{0})$ in the exotic SSPT phase and $\hat{C}^k$, $\hat{c}_{12}$ as
\ali{
    \hat{C}^{12}_0 &\simeq \hat{C}^1_0 - \hat{C}^2_0 \,,\label{3cor6} \\
    \hat{C} &\simeq \partial_1 \hat{C}^1_2 - \partial_2 \hat{C}^2_1 + \hat{c}_{12} \,, \label{3cor7}\\
    \hat{C}^{12}_3 &\simeq \hat{C}^1_3 - \hat{C}^2_3  \label{3cor8}\,,
}
where $\hat{C}^k$ is a $U(1)$ foliated $B$-type one-form gauge field and $\hat{c}_{12} \, dx^1 dx^2$ is a $U(1)$ bulk two-form gauge field. Their background gauge transformations are
\ali{
    \hat{C}^k &\rightarrow \hat{C}^k + d \hat{\lambda}^k - \hat{\nu}  \,, \\
    \hat{c}_{12} &\rightarrow \hat{c}_{12} + (d\hat{\nu})_{12} \,,
}
where $\hat{\lambda}^k$ is a zero-form gauge parameter and $\hat{\nu}$ is a one-form gauge parameter. Similarly, we can get correspondences between background gauge parameters:
\ali{
    \hat{\lambda}^{12} \simeq \hat{\lambda}^1 - \hat{\lambda}^2 \,.
}

To restrict $C$ and $\hat{C}$ fields to $\Z_N$ in the exotic SSPT phase, we have introduced the dynamical gauge fields $\hat{\beta}^{12}$ and $(\beta_0, \beta_{12}, \beta_3)$. We also introduce dynamical gauge fields $\hat{\beta}^k$, $\beta^k \wedge dx^k$, $\beta$, and assume correspondences 
\ali{
    \hat{\beta}^{12} &\simeq \hat{\beta}^1 - \hat{\beta}^2 \,, \\
    \partial_k \beta_0 &\simeq \beta^k_0 + \partial_0 \beta_k + c_{0k} \,, \quad (k=1,2) \,, \\
    \partial_k \beta_3 &\simeq \beta^k_3 + \partial_3 \beta_k + c_{3k} \,, \quad (k=1,2) \,, \\
    \beta_{12} &\simeq \beta^1_2 + \partial_2 \beta_1 = \beta^2_1 + \partial_1 \beta_2 + c_{12} \,, 
}
where $\hat{\beta}^k$ is a foliated $B$-type zero-form gauge field, $\beta^k\wedge dx^k$ is a foliated $A$-type (1+1)-form gauge field and $\beta$ is a bulk one-form gauge field. Moreover, to impose the constraint for $\beta^k\wedge dx^k$, $\beta$ and $c$, we introduce a bulk dynamical two-form gauge field $\hat{c}_{ij} dx^i dx^j \, ((i,j) = (0,1),(0,2),(0,3),(2,3),(3,1))$, and add the term $\frac{iN}{2\pi} \left( \sum_{k=1}^2 \beta^k \wedge dx^k + d\beta + c \right) \wedge \hat{c}\, |_{\hat{c}_{12} = 0}$ to the Lagrangian. Their background gauge transformations are
\ali{
    \hat{\beta}^k &\rightarrow \hat{\beta}^k + \hat{\lambda}^k  \,, \\
    \beta^k \wedge dx^k &\rightarrow \beta^k \wedge dx^k - \lambda^k \wedge dx^k  \,, \\ 
    \beta &\rightarrow \beta - \lambda  \,, \\
    \hat{c}_{ij} &\rightarrow \hat{c}_{ij} + (d\hat{\nu})_{ij} \,, \quad ( (i,j) \neq (1,2))  \,.
}
They also have dynamical gauge transformations:
\ali{
    \hat{\beta}^k &\rightarrow \hat{\beta}^k + 2\pi \hat{u}^k - \hat{v}   \,, \\
    \beta^k \wedge dx^k &\rightarrow \beta^k \wedge dx^k + d u^k \wedge dx^k   \,, \\
    \beta &\rightarrow \beta + d u - \sum_{k=1}^2 u^k dx^k \,,
}
where $\hat{u}^k$ is an $x^k$-dependent integer-valued gauge parameter, $u^k dx^k$ is a (0+1)-form gauge parameter, $s$ is a zero-form gauge parameter, and $\hat{v}$ is the gauge parameter of $p$. They correspond to the dynamical gauge parameters of $\hat{\beta}^{12}$ and $(\beta_0, \beta_{12}, \beta_3)$ as
\ali{
    \hat{s}^k &\simeq \hat{u}^k  \,, \\
    s &\simeq u \,.
}

Let us construct the foliated Lagrangian describing the 3+1d SSPT phase. The exotic SSPT phase with two foliations is described by the Lagrangian \eqref{sspte}
\sbali{
    &\L_{\text{SSPT},\text{e}} \left[ C_{012}, C_{312},C_{[03]}, \hat{C}^{12}_0, \hat{C}, \hat{C}^{12}_3 \right] = \L_{\text{SSPT},\text{e},\hat{\beta}} + \L_{\text{SSPT},\text{e},\beta} + \L_{\text{SSPT},\text{e}, C\hat{C}} \,, \\
    &\L_{\text{SSPT},\text{e}, \hat{\beta}} = \frac{iN}{2\pi} \hat{\beta}^{12} \left( \partial_3 C_{012} - \partial_0 C_{312} - \partial_1 \partial_2 C_{[03]} \right) \,, \\
    &\L_{\text{SSPT},\text{e},\beta} = \frac{iN}{2\pi} \left[ \beta_0 \left( \partial_3 \hat{C} - \partial_1 \partial_2 \hat{C}^{12}_3  \right) + \beta_{12} \left( \partial_3 \hat{C}^{12}_0 - \partial_0 \hat{C}^{12}_3  \right)  - \beta_3 \left( \partial_0 \hat{C} - \partial_1 \partial_2 \hat{C}^{12}_0  \right) \right]  \,, \\
    &\L_{\text{SSPT},\text{e}, C\hat{C}} =   \frac{iN}{2\pi}  \left( C_{012} \hat{C}^{12}_3 - C_{312} \hat{C}^{12}_0 + C_{[03]} \hat{C}  \right) \,.
}
Using the correspondences above, we can rewrite it in terms of the foliated fields. As for $\L_{\text{SSPT},\text{e}, \hat{\beta}}$ and $\L_{\text{SSPT},\text{e},\beta}$, we have
\alis{
    \L_{\text{SSPT},\text{e}, \hat{\beta}} \simeq & \frac{iN}{2\pi} \hat{\beta}^1 \left[ \partial_3 ( C^1_{02} + \partial_2 c_{01} ) - \partial_0 ( C^1_{32} + \partial_2 c_{31} ) - \partial_2 ( C^1_{03} + \partial_3 c_{01} - \partial_0 c_{31} ) \right] \\
    &- \frac{iN}{2\pi} \hat{\beta}^2 \left[ \partial_3 ( C^2_{01} + \partial_1 c_{02} - \partial_0 c_{12} ) - \partial_0 ( C^2_{31} + \partial_1 c_{32} - \partial_3 c_{12} ) \right. \\
    & \qquad \left. - \partial_1  ( C^2_{03} + \partial_3 c_{02} - \partial_0 c_{32} ) \right] \\
    = & \frac{iN}{2\pi} \sum_{k=1}^2 \hat{\beta}^k\, d C^k \wedge dx^k \,,
}
and
\alis{
    &\L_{\text{SSPT},\text{e},\beta} \\
    & \quad \simeq  \frac{iN}{2\pi} \left[ \beta_0 \left( \partial_3 \partial_1 \hat{C}^1_2  - \partial_3 \partial_2 \hat{C}^2_1 + \partial_3 \hat{c}_{12} - \partial_1 \partial_2 \hat{C}^1_3 + \partial_1 \partial_2 \hat{C}^2_3  \right) \right.   \\
    & \qquad + \beta_{12} \left( \partial_3 \hat{C}^1_0 - \partial_3 \hat{C}^2_0 - \partial_0 \hat{C}^1_3 + \partial_0 \hat{C}^2_3  \right) \\
    &\qquad \left. - \beta_3 \left( \partial_0 \partial_1 \hat{C}^1_2 - \partial_0 \partial_2 \hat{C}^2_1 + \partial_0 \hat{c}_{12} - \partial_1 \partial_2 \hat{C}^1_0 + \partial_1 \partial_2 \hat{C}^2_0  \right) \right] \\
    &\quad \simeq  \frac{iN}{2\pi} \left[ -\left(\beta^1_0 + \partial_0 \beta_1 + c_{01} \right) \left( \partial_3 \hat{C}^1_2 - \partial_2 \hat{C}^1_3 \right) -\left(\beta^2_0 + \partial_0 \beta_2 + c_{02} \right) \left( -\partial_3 \hat{C}^2_1  + \partial_1 \hat{C}^2_3  \right)  \right.   \\
    &\qquad + \beta_0 \partial_3 \hat{c}_{12} + \left( \beta^1_2 + \partial_2 \beta_1 \right) \left( \partial_3 \hat{C}^1_0 - \partial_0 \hat{C}^1_3 \right) + \left( \beta^2_1 + \partial_1 \beta_2 + c_{12} \right)\left( - \partial_3 \hat{C}^2_0  + \partial_0 \hat{C}^2_3  \right) \\
    &\qquad\left. + \left( \beta^1_3 + \partial_3 \beta_1 + c_{31} \right) \left( \partial_0 \hat{C}^1_2  - \partial_2 \hat{C}^1_0 \right) + \left( \beta^2_3 + \partial_3 \beta_2 + c_{32} \right) \left( - \partial_0 \hat{C}^2_1 + \partial_1 \hat{C}^2_0  \right) - \beta_3 \partial_0 \hat{c}_{12} \right] \\
    &\quad=  \frac{iN}{2\pi} \left[ \sum_{k=1}^2 \beta^k \wedge  dx^k \wedge d \hat{C}^k +  ( \beta_0 \partial_3 \hat{c}_{12} - \beta_3 \partial_0 \hat{c}_{12} ) d^4 x \right] + d \left[ \frac{iN}{2\pi} \sum_{k=1}^2 ( \beta_k \, dx^k )\wedge d\hat{C}^k  \right] \\
    &\qquad + \frac{iN}{2\pi} \left[ - c_{01} ( \partial_3 \hat{C}^1_2 - \partial_2 \hat{C}^1_3 ) + c_{02} ( \partial_3 \hat{C}^2_1 - \partial_1 \hat{C}^2_3 ) - c_{12} ( \partial_3 \hat{C}^2_0 - \partial_0 \hat{C}^2_3 ) \right. \\
    &\qquad \quad   \left. + c_{31} ( \partial_0 \hat{C}^1_2 - \partial_2 \hat{C}^1_0 ) - c_{32} ( \partial_0 \hat{C}^2_1 - \partial_1 \hat{C}^2_0 ) \right] d^4 x \,,
}
where $d^4 x = dx^0 dx^1 dx^2 dx^3$. For later convenience, we have left the total derivative term. For the constraints for $C^k \wedge dx^k$ and $c$, adding the term
\alis{
&\frac{iN}{2\pi} \left( \sum_{k=1}^2 \beta^k \wedge dx^k + d\beta + c \right) \wedge \hat{c}\, |_{\hat{c}_{12} = 0} \\
& \quad = \frac{iN}{2\pi}  \sum_{k=1}^2 \beta^k \wedge dx^k  \wedge \hat{c} + \left. \frac{iN}{2\pi}  \left[ \beta \wedge d \hat{c} + c \wedge \hat{c} + d ( \beta \wedge  \hat{c}) \right]\right|_{\hat{c}_{12} = 0} \,,
}
we have 
\alis{
    &\L_{\text{SSPT},\text{f},\beta} \\
    &\quad =  \frac{iN}{2\pi} \left[ \sum_{k=1}^2 \beta^k \wedge  dx^k \wedge ( d \hat{C}^k + \hat{c} ) + \beta \wedge d \hat{c} + ( c \wedge \hat{c})|_{\hat{c}_{12} = 0}  \right] \\
    &\qquad  + d \left[ \frac{iN}{2\pi} \sum_{k=1}^2 ( \beta_k \, dx^k )\wedge ( d\hat{C}^k + \hat{c} )  \right] \\
    &\qquad + \frac{iN}{2\pi} \left[ - c_{01} ( \partial_3 \hat{C}^1_2 - \partial_2 \hat{C}^1_3 ) + c_{02} ( \partial_3 \hat{C}^2_1 - \partial_1 \hat{C}^2_3 ) - c_{12} ( \partial_3 \hat{C}^2_0 - \partial_0 \hat{C}^2_3 ) \right. \\
    &\qquad \quad   \left. + c_{31} ( \partial_0 \hat{C}^1_2 - \partial_2 \hat{C}^1_0 ) - c_{32} ( \partial_0 \hat{C}^2_1 - \partial_1 \hat{C}^2_0 ) \right] \,, \label{lagbetac}
}
where we dropped the $x^0$-, $x^1$- and $x^2$-derivative term. Note that we combine the background gauge field $\hat{c}_{12}$ and the dynamical gauge fields $\hat{c}_{ij}$ $((i,j) = (0,1), (0,2), (0,3), (2,3), (3,1))$ into $\hat{c}$. To obtain the exotic theory, we must integrate out the dynamical parts $\hat{c}_{ij} \, ((i,j) = (0,1),(0,2),(0,3),(2,3),(3,1))$.

The $C\hat{C}$ part is rewritten as
\alis{
    \L_{\text{SSPT},\text{e}, C\hat{C}} \simeq &   \frac{iN}{2\pi}  \left[ ( C^1_{02} + \partial_2 c_{01} )\hat{C}^1_3 - ( C^2_{01} + \partial_1 c_{02} - \partial_0 c_{12}) \hat{C}^2_3 \right. \\
    & \left. - ( C^1_{32} + \partial_2 c_{31} )  \hat{C}^1_0 + ( C^2_{31} + \partial_1 c_{32} - \partial_3 c_{12} ) \hat{C}^2_0  + c_{03} (\partial_1 \hat{C}^1_2 - \partial_2 \hat{C}^2_1 + \hat{c}_{12} )  \right] \\
    = & \frac{iN}{2\pi} \left[ \sum_{k=1}^2 \hat{C}^k \wedge  C^k \wedge dx^k + \hat{c}_{12} c_{03}\,  d^4x \right] \\
    & + \frac{iN}{2\pi} \left[ \partial_2 c_{01}\, \hat{C}^1_3 
    - (\partial_1 c_{02} - \partial_0 c_{12}) \hat{C}^2_3 - \partial_2 c_{31} \, \hat{C}^1_0 + ( \partial_1 c_{32} - \partial_3 c_{12} ) \hat{C}^2_0  \right. \\ 
    & \left. \quad - ( \partial_3 c_{01} - \partial_0 c_{31} ) \hat{C}^1_2 + ( \partial_3 c_{02} - \partial_0 c_{32} ) \hat{C}^2_1   \right] d^4 x \,.
}
Then, combining the last term with the last term of \eqref{lagbetac}, we have
\alis{
    \frac{iN}{2\pi} \left[ \partial_3 \left( - c_{01} \hat{C}^1_2 + c_{02} \hat{C}^2_1 - c_{12} \hat{C}^2_0   \right) \right] d^4 x \,. \label{bdcc}
}
We will put the SSPT on the region $x^3 \geq 0$ and the foliated $BF$ theory on the boundary $x^3 = 0$. Then, the $x^3$-derivative term cancels out the boundary term \eqref{dropterm}, so we drop this term.

Finally, we obtain the foliated form of the SSPT phase with two foliations in 3+1 dimensions:
\alis{
    &\L_{\text{SSPT},\text{f}} \left[ C^k \wedge dx^k , \hat{C}^k, c, \hat{c}_{12} \right]  \\
    &\qquad   = \frac{iN}{2\pi}  \left\{ \sum_{k=1}^2 \left[ \hat{\beta}^k\, d C^k \wedge dx^k
    + \beta^k \wedge  dx^k \wedge  ( d \hat{C}^k + \hat{c} ) \right] +  \beta \wedge d\hat{c} \right\} \\
     &\quad \qquad + d \left[ \frac{iN}{2\pi} \sum_{k=1}^2 (\beta_k \, dx^k) \wedge ( d\hat{C}^k + \hat{c} ) \right] + \frac{iN}{2\pi} \left[ \sum_{k=1}^2 \hat{C}^k \wedge  C^k \wedge dx^k + c \wedge \hat{c} \right] \\ 
     &\quad \qquad  + \frac{iN}{2\pi} p \wedge  \left( \sum_{k=1}^2 C^k \wedge dx^k - dc \right)  \,. \label{folisspt}
}

As in the case of the exotic SSPT phase, if the theory is on spacetime without a boundary, it is gauge invariant. However if spacetime has a boundary, the partition function of the 3+1d SSPT phase is not invariant. From the anomaly inflow mechanism \cite{Callan:1984sa}, this variation is expected to be canceled by the anomaly of the 2+1d foliated $BF$ theory on the boundary \eqref{fanomaly}. To see this, we put the foliated SSPT phase on the region $x^3 \geq 0$ with the boundary $x^3 = 0$. From the gauge invariance,  the boundary conditions of $\hat{\beta}^k$, $\beta^k \wedge dx^k$ and $\beta$ are
\ali{
    ( \hat{\beta}^1 - \hat{\beta}^2 )  |_{x^3 = 0} &= 0 \,, \\ 
    \beta_0\, |_{x^3 = 0} &= 0 \,, \\
    (\beta^1_2 + \partial_2 \beta_1) |_{x^3 = 0}  &= 0 \,, \\
    (\beta^2_1 + \partial_1 \beta_2 + c_{12} ) |_{x^3 = 0} &= 0 \,,
}
which are consistent with the boundary conditions in the exotic SSPT phase. On the boundary, we put the 2+1d foliated $BF$ theory coupled to the background gauge fields \eqref{fbfc}, and the background gauge fields in the 3+1d SSPT phase are related to those in the 2+1d foliated $BF$ theory as\footnote{On the boundary, $\hat{c}_{\text{SSPT},01}$ and $\hat{c}_{\text{SSPT},02}$ do not arise. Then, these have no relation to $\hat{c}_{BF,01}$ and $\hat{c}_{BF,02}$.}
\ali{
    C^k_{\text{SSPT}} \wedge dx^k \, |_{x^3 = 0} &= C^k_{BF} \wedge dx^k \,, \\
    \hat{C}^k_{\text{SSPT}}\, |_{x^3 = 0} &= \hat{C}^k_{BF} \,, \\ 
    c_{\text{SSPT}}\, |_{x^3 = 0} &= c_{BF} \,, \\
    \hat{c}_{\text{SSPT},12}\, |_{x^3 = 0} &= \hat{c}_{BF,12} \,.
}
Note that while the background gauge fields in the 3+1d SSPT phase are restricted to $\Z_N$ tensor gauge fields by the dynamical fields $\hat{\beta}^k$, $\beta^k \wedge dx^k$ and $\beta$, those in the 2+1d foliated $BF$ theory are restricted by the dynamical fields $\hat{\chi}^k$, $\chi^k \, dx^k$ and $\chi$.
Then, under the background gauge transformations, the Lagrangian transforms as
\alis{
    \delta_g \L_{\text{SSPT,f}} &=  \frac{iN}{2\pi} \sum_{k=1}^2 \left[  \hat{\lambda}^k\, d C^k  \wedge dx^k   -\lambda^k \wedge  dx^k \wedge  (d \hat{C}^k + \hat{c} ) \right] - \frac{iN}{2\pi}  \lambda \wedge d\hat{c} \\
    & - d \left[ \frac{iN}{2\pi} \sum_{k=1}^2 (\lambda_k \, dx^k) \wedge ( d\hat{C}^k + \hat{c} ) \right] \\
    &+ \frac{iN}{2\pi} \sum_{k=1}^2 \left[  (d\hat{\lambda}^k - \hat{\nu} ) \wedge   C^k \wedge dx^k 
     +  ( \hat{C}^k + d\hat{\lambda}^k - \hat{\nu} ) \wedge d\lambda^k \wedge dx^k \right. \\
    &+ \left. \left( d\lambda + \sum^2_{k=1} \lambda^k \wedge dx^k \right) \wedge \hat{c} + \left( c + d\lambda + \sum^2_{k=1} \lambda^k \wedge dx^k \right) \wedge d\hat{\nu}  \right] \\
    &= d \left\{ \frac{iN}{2\pi} \sum_{k=1}^2  \left[  \hat{\lambda}^k\, C^k  \wedge dx^k   - (\hat{C}^k + d\hat{\lambda}^k ) \wedge (\lambda^k \wedge  dx^k + d(\lambda_k\, dx^k)  \right] \right. \\
    & \left. \quad + \frac{iN}{2\pi} \lambda_0  \hat{c}_{12} d^3 x + \frac{iN}{2\pi} \hat{\nu} \wedge \left( c + d\lambda + \sum^2_{k=1} \lambda^k \wedge dx^k \right) \right\} \,,
}
Thus on the boundary, the term 
\alis{
    \delta_g S_{\text{SSPT,e}} &= -\int dx^0 dx^1 dx^2 \frac{iN}{2\pi}  \left\{  \sum_{k=1}^2  \left[  \hat{\lambda}^k\, C^k  \wedge dx^k   - (\hat{C}^k + d\hat{\lambda}^k ) \wedge (\lambda^k \wedge  dx^k + d(\lambda_k\, dx^k)  \right] \right. \\
    & \quad \qquad \left. + \lambda_0  \hat{c}_{12} d^3 x + \hat{\nu} \wedge \left( c + d\lambda + \sum^2_{k=1} \lambda^k \wedge dx^k \right)\right\} \label{anomcan2}
}
arises. Note the total derivative term in \eqref{folisspt} contribute to the boundary $x^3 = 0$ as a counterterm, so that the boundary term of the 3+1d foliated SSPT phase matches the 't Hooft anomaly of the 2+1d foliated $BF$ theory \eqref{fanomaly} on the boundary $x^3=0$. Therefore we can cancel the 't Hooft anomaly of the 2+1d foliated $BF$ theory on the boundary by the gauge-variation of the 3+1d foliated SSPT phase on the bulk.

\section{Change of Foliation Structure}
\label{section 4}

In Section \ref{section 33}, we obtained the 3+1d foliated SSPT phase with two foliations:
\alis{
    &\L^2_{\text{SSPT},\text{f}} \left[ C^k \wedge dx^k , \hat{C}^k, c, \hat{c}_{12} \right]  \\
    &\qquad   = \frac{iN}{2\pi}  \left\{ \sum_{k=1}^2 \left[ \hat{\beta}^k\, d C^k \wedge dx^k
    + \beta^k \wedge  dx^k \wedge  ( d \hat{C}^k + \hat{c} ) \right] +  \beta \wedge d\hat{c} \right\} \\
     &\quad \qquad + d \left[ \frac{iN}{2\pi} \sum_{k=1}^2 (\beta_k \, dx^k) \wedge ( d\hat{C}^k + \hat{c} ) \right] + \frac{iN}{2\pi} \left[ \sum_{k=1}^2 \hat{C}^k \wedge  C^k \wedge dx^k + c \wedge \hat{c} \right] \\ 
     &\quad \qquad  + \frac{iN}{2\pi} p \wedge  \left( \sum_{k=1}^2 C^k \wedge dx^k - dc \right)  \,. \label{2folisspt4}
}

Here we change the foliation structure from two foliations $e^k = dx^k\, (k=1,2)$ to three foliations $e^k = dx^k\, (k=1,2,3)$. By adding gauge fields with $k=3$ and modifying the gauge transformations, we can easily construct the 3+1d foliated SSPT phase with three foliations. The SSPT phase with a boundary also cancels the mixed 't Hooft anomaly of the 2+1d foliated $BF$ theory on the boundary. The situation where other (exotic) SSPT phases with different foliations cancel the same anomaly appears in \cite{Burnell:2021reh,Luo:2022mrj}.

Then, we will assume field correspondences between the foliated and exotic SSPT phases with three foliations, and convert from the foliated SSPT phase with three foliations to the exotic one. While it is non-trivial to construct the exotic form of the 3+1d SSPT phase with three foliations from the one with two foliations, we can construct it via the foliated form. It is a systematic way to construct the exotic form using the foliated-exotic duality.

\subsection{Foliated SSPT phase with three foliations}
\label{section 41}

Firstly, we construct the 3+1d foliated SSPT phase with three foliations. We introduce a $U(1)$ background foliated $A$-type (2+1)-form gauge fields $C^k \wedge dx^k\, (k = 1,2,3)$ and a $U(1)$ background bulk two-form gauge field $c$ with background gauge transformations
\ali{
    C^k \wedge dx^k &\rightarrow C^k \wedge dx^k + d\lambda^k \wedge dx^k \,, \quad (k=1,2,3) \,, \\
    c &\rightarrow c + d\lambda + \sum_{k=1}^3 \lambda^k \wedge dx^k \,,
}
where $\lambda^k \wedge dx^k$ is a (1+1)-form gauge parameter and $\lambda$ is a one-form gauge parameter. As in the case of two foliations, to impose the constraints for $C^k \wedge dx^k$ and $c$, we include the term $\frac{iN}{2\pi} p \wedge  \left(\sum_{k=1}^3 C^k \wedge dx^k - dc \right)$ in the Lagrangian, where $p$ is a dynamical one-form field. $p$ has a dynamical gauge transformation $p \rightarrow p + d\hat{v}$, where $\hat{v}$ is a zero-form gauge parameter. Next, we introduce $U(1)$ background foliated $B$-type one-form gauge fields $\hat{C}^k\, (k = 1,2,3)$ and a $U(1)$ background bulk gauge field $\hat{c}_{ij} \, dx^i dx^j \, ((i,j) = (1,2), (2,3), (3,1))$ with background gauge transformations
\ali{
    \hat{C}^k &\rightarrow \hat{C}^k + d \hat{\lambda}^k - \hat{\nu}  \,, \\
    \hat{c}_{ij} &\rightarrow \hat{c}_{ij} + (d\hat{\nu})_{ij} \,, \quad ((i,j) = (1,2), (2,3), (3,1)) \,,
}
where $\hat{\lambda}^k$ is a zero-form gauge parameter and $\hat{\nu}$ is a one-form gauge parameter. To restrict $C^k \wedge dx^k$, $\hat{C}^k$ and $\hat{c}_{ij}\, ((i,j) = (1,2), (2,3), (3,1))$ to $\Z_N$, we introduce a dynamical foliated $B$-type zero-form gauge fields $\hat{\beta}^k \, (k =1,2,3)$, a dynamical foliated $A$-type (1+1)-form gauge fields $\beta^k \wedge dx^k  \, (k =1,2,3)$ and a dynamical bulk one-form gauge field $\beta$. Moreover, to impose the constraint for $\beta^k\wedge dx^k$ and $\beta$, we introduce a bulk dynamical two-form gauge field $\hat{c}_{ij} dx^i dx^j \, ((i,j) = (0,1),(0,2),(0,3))$, and include the term $\frac{iN}{2\pi} \left( \sum_{k=1}^3 \beta^k \wedge dx^k + d\beta  + c \right) \wedge \hat{c}\, |_{\hat{c}_{12} = \hat{c}_{23} = \hat{c}_{31} = 0}$ in the Lagrangian. Their background gauge transformations are
\ali{
    \hat{\beta}^k &\rightarrow \hat{\beta}^k + \hat{\lambda}^k  \,, \\
    \beta^k \wedge dx^k &\rightarrow \beta^k \wedge dx^k - \lambda^k \wedge dx^k  \,, \\ 
    \beta &\rightarrow \beta - \lambda  \,, \\
    \hat{c}_{ij} &\rightarrow \hat{c}_{ij} + (d\hat{\nu})_{ij} \,, \quad  ((i,j) = (0,1),(0,2),(0,3))  \,.
}
They also have dynamical gauge transformations:
\ali{
    \hat{\beta}^k &\rightarrow \hat{\beta}^k + 2\pi \hat{u}^k - \hat{v}   \,, \\
    \beta^k \wedge dx^k &\rightarrow \beta^k \wedge dx^k + d u^k \wedge dx^k   \,, \\
    \beta &\rightarrow \beta + d u - \sum_{k=1}^3 u^k dx^k \,,
}
where $\hat{u}^k$ is a $x^k$-dependent integer-valued gauge parameter, $u^k dx^k$ is a (0+1)-form gauge parameter, $s$ is a zero-form gauge parameter, and $\hat{v}$ is the gauge parameter of $p$. Then, the 3+1d foliated SSPT phase with three foliations is written as
\alis{
    &\L^3_{\text{SSPT},\text{f}} \left[ C^k \wedge dx^k,  \hat{C}^k, c, \hat{c}_{12}, \hat{c}_{23}, \hat{c}_{31} \right]  \\
    &\qquad   = \frac{iN}{2\pi}  \left\{ \sum_{k=1}^3 \left[ \hat{\beta}^k\, d C^k \wedge dx^k
    + \beta^k \wedge  dx^k \wedge  ( d \hat{C}^k + \hat{c} ) \right] +  \beta \wedge d\hat{c} \right\} \\
     &\quad \qquad + d \left[ \frac{iN}{2\pi} \sum_{k=1}^3 (\beta_k \, dx^k) \wedge ( d\hat{C}^k + \hat{c} ) \right] + \frac{iN}{2\pi}  \left( \sum_{k=1}^3 \hat{C}^k \wedge  C^k \wedge dx^k + c \wedge \hat{c} \right) \\ 
     &\quad \qquad  + \frac{iN}{2\pi} p \wedge   \left( \sum_{k=1}^3 C^k \wedge dx^k - dc \right)   \,. \label{3folisspt4}
}

If the theory is on spacetime without a boundary, it is gauge invariant. As in the case of the SSPT phase with two foliations, if spacetime has a boundary, the partition function of the 3+1d SSPT phase with three foliations is not gauge invariant and the variation is canceled by the anomaly of the 2+1d foliated $BF$ theory on the boundary \eqref{fanomaly}. To see this, we put the foliated SSPT phase on the region $x^3 \geq 0$ with the boundary $x^3 = 0$. From the gauge invariance, the boundary conditions of $\hat{\beta}^k$, $\beta^k \wedge dx^k$ and $\beta$ are
\ali{
    (\hat{\beta}^1 - \hat{\beta}^2 ) |_{x^3 = 0} &= 0 \,, \\ 
    \beta_0\, |_{x^3 = 0} &= 0 \,, \\
    (\beta^1_2 + \partial_2 \beta_1 ) |_{x^3 = 0}  &= 0 \,, \\
    (\beta^2_1 + \partial_1 \beta_1 + c_{12} ) |_{x^3 = 0} &= 0 \,.
}
On the boundary, we put the 2+1d foliated $BF$ theory coupled to the background gauge fields \eqref{fbfc}, and the background gauge fields in the 3+1d SSPT phase are related to those in the 2+1d foliated $BF$ theory as
\ali{
    C^k_{\text{SSPT}} \wedge dx^k \, |_{x^3 = 0} &= C^k_{BF} \wedge dx^k \,, \quad (k=1,2) \,, \\
    \hat{C}^k_{\text{SSPT}}\, |_{x^3 = 0} &= \hat{C}^k_{BF} \,, \quad (k=1,2) \,, \\
    c_{\text{SSPT}} \, |_{x^3 = 0} &= c_{BF} \,, \\
    \hat{c}_{\text{SSPT},12}\, |_{x^3 = 0} &= \hat{c}_{BF,12} \,.
}
Then, under the background gauge transformations, the Lagrangian transforms as
\alis{
    \delta_g \L^3_{\text{SSPT,f}} &=  \frac{iN}{2\pi} \sum_{k=1}^3 \left[  \hat{\lambda}^k\, d C^k  \wedge dx^k   -\lambda^k \wedge  dx^k \wedge  (d \hat{C}^k + \hat{c} ) \right] - \frac{iN}{2\pi}  \lambda \wedge d\hat{c} \\
    &\quad  - d \left[ \frac{iN}{2\pi} \sum_{k=1}^3 ( \lambda_k \, dx^k) \wedge ( d\hat{C}^k + \hat{c} ) \right] \\
    &\quad + \frac{iN}{2\pi} \sum_{k=1}^3 \left[  (d\hat{\lambda}^k - \hat{\nu} ) \wedge   C^k \wedge dx^k +  ( \hat{C}^k + d\hat{\lambda}^k - \hat{\nu} ) \wedge d\lambda^k \wedge dx^k \right] \\ 
    &\quad +  \frac{iN}{2\pi} \left( d\lambda + \sum^3_{k=1} \lambda^k \wedge dx^k \right) \wedge \hat{c} + \frac{iN}{2\pi} \left( c + d\lambda + \sum^3_{k=1} \lambda^k \wedge dx^k \right) \wedge d\hat{\nu}    \\
    &= d \left\{ \frac{iN}{2\pi} \sum_{k=1}^3  \left[  \hat{\lambda}^k\, C^k  \wedge dx^k   - (\hat{C}^k + d\hat{\lambda}^k ) \wedge (\lambda^k \wedge  dx^k + d(\lambda_k\, dx^k)  \right] \right. \\
    & \quad \left. + \frac{iN}{2\pi} \lambda_0  \hat{c}_{12} d^3 x + \hat{\nu} \wedge  \left( c + d\lambda + \sum^3_{k=1} \lambda^k \wedge dx^k \right) \right\} \,.
}
On the boundary $x^3 = 0$, the terms containing $dx^3$, such as the foliated fields with $k = 3$, do not appear, so we have 
\alis{
    \delta_g S^3_{\text{SSPT,e}} &= -\int dx^0 dx^1 dx^2 \frac{iN}{2\pi}  \left\{  \sum_{k=1}^2  \left[  \hat{\lambda}^k\, C^k  \wedge dx^k   - (\hat{C}^k + d\hat{\lambda}^k ) \wedge (\lambda^k \wedge  dx^k + d(\lambda_k\, dx^k)  \right] \right. \\
    & \qquad \quad \left. + \lambda_0  \hat{c}_{12} d^3 x  + \hat{\nu} \wedge  \left( c + d\lambda + \sum^3_{k=1} \lambda^k \wedge dx^k \right) \right\} \,.
}
This expression is the same as the boundary term of the 3+1d foliated SSPT with two foliations \eqref{anomcan2}, and matches the 't Hooft anomaly of the 2+1d foliated $BF$ theory \eqref{fanomaly} on the boundary $x^3=0$. Therefore we can also cancel the 't Hooft anomaly of the 2+1d foliated $BF$ theory on the boundary by the gauge variation of the 3+1d foliated SSPT phase with three foliations on the bulk. 

An 't Hooft anomaly of an ordinary global symmetry in a relativistic QFT corresponds to a field theory in one dimension higher from the anomaly inflow \cite{Callan:1984sa}, and such field theories are called invertible field theories \cite{Freed:2016rqq}. Invertible field theories are the low-energy effective field theories of the symmetry-protected topological (SPT) phases \cite{Gu:2009dr,Chen:2010gda}, which cannot be smoothly deformed into trivially gapped systems while preserving the symmetry.

In the case of the subsystem symmetries, the 3+1d SSPT phases with two foliations and three foliations represent the same phases from an anomaly point of view. The rotational symmetry of the SSPT phase with two foliations is the 90 degree rotation $\Z_4$ with respect to $(x^1,x^2)$ and the $x^3$ direction has no foliation. On the other hand, the rotational symmetry of the SSPT phase with three foliations is the 90 degree rotation $S_4$ with respect to $(x^1,x^2,x^3)$. On the boundary $x^3 =0$, both reproduce the 90 degree rotational symmetry $\Z_4$ in the exotic/foliated $BF$ theory in 2+1 dimensions. Furthermore, the SSPT phase with three foliations \eqref{3folisspt4} is smoothly deformed into the one with two foliations \eqref{2folisspt4} while preserving the rotational symmetry $\Z_4$ under the deformation $e^3 = dx^3 \rightarrow 0$, that is $\hat{\beta}^3$, $\beta^3 \wedge dx^3$, $C^3 \wedge dx^3$ and $\hat{C}^3$ go to $0$.\footnote{Precisely, we must consider the version where the bulk gauge fields $\hat{c}$ are not dynamical. This fact implies that we cannot apply the same discussion to the exotic SSPT phases.} Therefore, an 't Hooft anomaly of a subsystem symmetry in a fractonic QFT is considered to correspond to a certain deformation class of the SSPT phases and foliation structures. This gives an implication for the characterization of 't Hooft anomalies of subsystem symmetry.

\subsection{Exotic SSPT phase with three foliations}
\label{section 42}

In Section \ref{section 41}, we derived the foliated SSPT phase with three foliations simply by adding the $k = 3$ foliation terms. Here assuming field correspondences, we determine tensor gauge fields of the SSPT phase with three foliations and construct the exotic Lagrangian describing the SSPT phase with three foliations.

To derive the exotic form, we must integrate out $\hat{c}_{01}$, $\hat{c}_{02}$, $\hat{c}_{03}$ and $p$ in the foliated form, and then we can use the equations of motion
\ali{
    \frac{iN}{2\pi} (\beta^i_j  - \beta^j_i + \partial_j \beta_i - \partial_i \beta_j - c_{ij} ) &= 0 \,, \quad ((i,j) = (1,2), (2,3), (3,1)) \,, \\
    \frac{iN}{2\pi} \left(\sum_{k=1}^3 C^k \wedge dx^k - dc \right) &= 0 \,. \label{3peom3}
}

The fractonic theory with three simultaneous foliations $e^k = dx^k\, (k=1,2,3)$ has the 90 degree rotational symmetry $S_4$ in 3+1 dimensions. Then, tensor gauge fields of such a theory are in representations of $S_4$. Irreducible representations of $S_4$ are $\bm{1}$, $\bm{1}'$, $\bm{2}$,  $\bm{3}$ and $\bm{3}'$, and we use the notation in \cite{Seiberg:2020wsg}. Firstly, we introduce $U(1)$ background tensor gauge fields $(C_{012}, C_{023}, C_{031})$ in $\bm{3}'$ and $(C_{1(23)},C_{2(31)},C_{3(12)})$ in $\bm{2}$ satisfying $C_{1(23)} + C_{2(31)} + C_{3(12)} =0$, and assume correspondences
\ali{
    C_{0ij} &\simeq C^i_{0j} + \partial_j c_{0i} = C^j_{0i} + \partial_i c_{0j} - \partial_0 c_{ij} \,, \quad ((i,j) = (1,2), (2,3), (3,1)) \,, \label{33cor1} \\
    C_{i(jk)} &\simeq C^k_{ij} - C^j_{ki}  - \partial_i c_{jk} + \partial_k c_{ij} \,, \quad ((i,j,k) = (1,2,3),(2,3,1),(3,1,2))  \,, \label{33cor2} 
}
or\footnote{We can use $(C_{[12]3},C_{[23]1},C_{[31]2})$ in $\bm{2}$ satisfying $C_{[ij]k} = \frac{1}{3} \left( C_{i(jk)} - C_{j(ki)} \right)$ and $C_{i(jk)} = C_{[ij]k} - C_{[ki]j} $.}
\ali{
\frac{1}{3} \left( C_{i(jk)} - C_{j(ki)} \right) \simeq C^k_{ij} - \partial_i c_{jk} \,, \quad ((i,j,k) = (1,2,3),(2,3,1),(3,1,2))  \,, \label{33cor2'} 
}
where we have used \eqref{3peom3} and actually we have restricted $c$ to zero. Their background gauge transformations are
\ali{
    C_{0ij} &\rightarrow C_{0ij} + \partial_0 \Lambda_{ij} - \partial_i \partial_j \Lambda_0 \,, \quad ((i,j) = (1,2),(2,3),(3,1)) \,, \\
    C_{k(ij)} &\rightarrow C_{k(ij)} + 2 \partial_k \Lambda_{ij} - \partial_i \Lambda_{jk} - \partial_j \Lambda_{ki} \,, \quad ((i,j,k) = (1,2,3),(2,3,1),(3,1,2)) \,,
}
where the background gauge parameters $\Lambda_0$ and $(\Lambda_{12},\Lambda_{23},\Lambda_{31})$ are in the representation $\bm{1}$ and $\bm{3}'$ respectively. Then, we have correspondences
\ali{
    \Lambda_0 &\simeq \lambda_0 \,, \\
    \Lambda_{ij} &\simeq \lambda^i_j + \partial_j \lambda_i   \,, \quad ((i,j) = (1,2),(2,3),(3,1)) \,.
}
Next, we introduce $U(1)$ background tensor gauge fields $(\hat{C}_0^{1(23)}, \hat{C}_0^{2(31)}, \hat{C}_0^{3(12)})$ in $\bm{2}$ satisfying $\hat{C}^{1(23)} + \hat{C}^{2(31)} + \hat{C}^{3(12)} =0$, $(\hat{C}^{12}, \hat{C}^{23}, \hat{C}^{31})$ in $\bm{3}'$ and $(\hat{C}^1, \hat{C}^2, \hat{C}^3)$ in $\bm{3}$, and assume correspondences
\ali{
    \hat{C}_0^{k(ij)} &\simeq \hat{C}_0^i - \hat{C}_0^j \,, \quad ((i,j,k) = (1,2,3),(2,3,1),(3,1,2)) \,, \label{33cor3} \\
    \hat{C}^{ij} &\simeq \hat{C}_k^i - \hat{C}_k^j \,, \quad ((i,j,k) = (1,2,3),(2,3,1),(3,1,2))  \,, \label{33cor4} \\
    \hat{C}^k &\simeq \partial_i \hat{C}_j^i - \partial_j \hat{C}_i^j + \hat{c}_{ij} \,, \quad ((i,j,k) = (1,2,3),(2,3,1),(3,1,2)) \label{33cor5} \,.
}
Note that the gauge transformations $\hat{\nu}$ in the right hand sides cancel out, so the degrees of freedom of the fields are consistent. Their background gauge transformations are
\ali{
    \hat{C}_0^{k(ij)} &\rightarrow \hat{C}_0^{k(ij)} + \partial_0 \hat{\Lambda}^{k(ij)}  \,, \quad ((i,j,k) = (1,2,3),(2,3,1),(3,1,2)) \,, \\
    \hat{C}^{ij} &\rightarrow \hat{C}^{ij} + \partial_k \hat{\Lambda}^{k(ij)} \,, \quad ((i,j,k) = (1,2,3),(2,3,1),(3,1,2)) \,, \\
    \hat{C}^k &\rightarrow \hat{C}^k + \partial_i \partial_j \hat{\Lambda}^{k(ij)} \,, \quad ((i,j,k) = (1,2,3),(2,3,1),(3,1,2)) \,,
}
where the background gauge parameters $(\hat{\Lambda}^{1(23)}, \hat{\Lambda}^{2(31)}, \hat{\Lambda}^{3(12)})$ are in the representation $\bm{2}$ satisfying $\hat{\Lambda}^{1(23)} + \hat{\Lambda}^{2(31)} + \hat{\Lambda}^{3(12)} = 0$. Then, we have correspondences
\ali{
    \hat{\Lambda}^{k(ij)} &\simeq \hat{\lambda}^i - \hat{\lambda}^j \,, \quad ((i,j,k) = (1,2,3),(2,3,1),(3,1,2)) \,.
}
In addition, to restrict $C$ and $\hat{C}$ fields to $\Z_N$, we introduce dynamical gauge fields \\
\noindent $(\hat{\beta}^{1(23)}, \hat{\beta}^{2(31)}, \hat{\beta}^{3(12)})$ in $\bm{2}$ satisfying $\hat{\beta}^{1(23)} + \hat{\beta}^{2(31)} + \hat{\beta}^{3(12)} =0$, $(\beta_{01}, \beta_{02}, \beta_{03})$ in $\bm{3}$ and \\
\noindent $(\beta_{12}, \beta_{23}, \beta_{31})$ in $\bm{3}'$, and assume correspondences
\ali{
    \hat{\beta}^{k(ij)} &\simeq \hat{\beta}^i - \hat{\beta}^j \,, \quad ((i,j,k) = (1,2,3),(2,3,1),(3,1,2)) \,, \\
    \beta_{0k} &\simeq \beta^k_0 + \partial_0 \beta_k + c_{0k} \,, \quad (k=1,2,3) \,, \\
    \beta_{ij} &\simeq \beta^i_j + \partial_j \beta_i = \beta^j_i + \partial_i \beta_j + c_{ij} \,, \quad ((i,j) = (1,2),(2,3),(3,1)) \,,
}
and use $\beta_0$ as a field in the representation $\bm{1}$. Their background gauge transformations are
\ali{
    \hat{\beta}^{k(ij)} &\rightarrow \hat{\beta}^{k(ij)} + \hat{\Lambda}^{k(ij)} \,, \quad ((i,j,k) = (1,2,3),(2,3,1),(3,1,2)) \,, \\
    \beta_0 &\rightarrow \beta_0 - \Lambda_0 \,, \\
    \beta_{0k} &\rightarrow \beta_{0k} - \partial_k \Lambda_0 \,, \quad (k=1,2,3) \,, \\
    \beta_{ij} &\rightarrow \beta_{ij} - \Lambda_{ij} \,, \quad ((i,j) = (1,2),(2,3),(3,1)) \,.
}
They also have dynamical gauge transformations
\ali{
    \hat{\beta}^{k(ij)} &\rightarrow \hat{\beta}^{k(ij)} + 2\pi \hat{s}^i - 2\pi \hat{s}^j \,, \quad ((i,j,k) = (1,2,3),(2,3,1),(3,1,2)) \,, \\
    \beta_0 &\rightarrow \beta_0 + \partial_0 s \,, \\
    \beta_{0k} &\rightarrow \beta_{0k} + \partial_k \partial_0 s \,, \quad (k=1,2,3) \,, \\
    \beta_{ij} &\rightarrow \beta_{ij} + \partial_i \partial_j s \,, \quad ((i,j) = (1,2),(2,3),(3,1)) \,,
}
where $\hat{s}^k$ is a $x^k$-dependent integer-valued gauge parameter, and $s$ is a gauge
parameter in $\bm{1}$, and we have
\ali{
    \hat{s}^k &\simeq \hat{u}^k  \,, \\
    s &\simeq u \,.
}

Using these field correspondences, we construct the exotic Lagrangian describing the 3+1d SSPT phase with three foliations. As for the $\L^3_{\text{SSPT,f},\hat{\beta}}$ part, we have
\alis{
   \L^3_{\text{SSPT,f},\hat{\beta}} &= \frac{iN}{2\pi}   \sum_{k=1}^3  \hat{\beta}^k\, d C^k \wedge dx^k \\
   &\simeq \frac{iN}{2\pi} \hat{\beta}^1 \left[ \partial_3 C_{012} - \partial_2 C_{031} -  \frac{1}{3} \partial_0 (C_{3(21)} - C_{2(31)}) \right] \\
   & \quad - \frac{iN}{2\pi} \hat{\beta}^2 \left[ \partial_3 C_{012} - \partial_1 C_{023} - \frac{1}{3} \partial_0 (C_{3(12)} - C_{1(23)})  \right] \\
   & \quad - \frac{iN}{2\pi} \hat{\beta}^3 \left[  \partial_1 C_{023} - \partial_2 C_{031} - \frac{1}{3} \partial_0 (C_{1(23)} - C_{2(31)})  \right] \\
   &\simeq \frac{iN}{2\pi} \sum_{(i,j,k)} \hat{\beta}^{k(ij)} \left( \partial_k C_{0ij} - \frac{1}{3} \partial_0 C_{k(ij)} \right) \,,
}
where the sum for $(i,j,k)$ is over $(i,j,k) = (1,2,3),(2,3,1),(3,1,2)$. Note that terms of the gauge field $c$ cancel out. As for the $\L^3_{\text{SSPT,f},\beta}$ part, we have
\alis{
    & \L^3_{\text{SSPT,f},\beta} \\
    &\quad = \frac{iN}{2\pi}  \left[ \sum_{k=1}^3  \beta^k \wedge  dx^k \wedge  ( d \hat{C}^k + \hat{c} ) +  \beta \wedge d\hat{c} \right]
      + d \left[ \frac{iN}{2\pi} \sum_{k=1}^3 (\beta_k \, dx^k) \wedge ( d\hat{C}^k + \hat{c} ) \right] \\
      & \quad= \frac{iN}{2\pi}  \left[ (\beta^1_2 + \partial_2 \beta_1) ( \partial_3 \hat{C}^1_0 - \partial_0 \hat{C}^1_3 )  + (\beta^2_1+ \partial_1 \beta_2 + c_{12}) ( - \partial_3 \hat{C}^2_0 + \partial_0 \hat{C}^2_3 ) \right. \\
      &\quad  \quad  \left. + (\beta^2_3+ \partial_3 \beta_2) ( - \partial_0 \hat{C}^2_1 + \partial_1 \hat{C}^2_0 ) + (\beta^3_2 + \partial_2 \beta_3 + c_{23}) (  \partial_0 \hat{C}^3_1 - \partial_1 \hat{C}^3_0 )  \right. \\
      &\quad \quad \left. + (\beta^3_1 + \partial_1 \beta_3) ( - \partial_0 \hat{C}^3_2 + \partial_2 \hat{C}^3_0 ) + (\beta^1_3 + \partial_3 \beta_1 + c_{31}) (  \partial_0 \hat{C}^1_2 - \partial_2 \hat{C}^1_0 )  \right. \\
      &\quad \quad \left. + (\beta^1_0 + \partial_0 \beta_1 + c_{01})( \partial_2 \hat{C}^1_3 - \partial_3 \hat{C}^1_2 + \hat{c}_{23}) + (\beta^2_0 + \partial_0 \beta_2 + c_{02})( \partial_3 \hat{C}^2_1 - \partial_1 \hat{C}^2_3 + \hat{c}_{31})  \right. \\
      &\quad \quad \left. + (\beta^3_0 + \partial_0 \beta_3 + c_{03})( \partial_1 \hat{C}^3_2 - \partial_2 \hat{C}^3_1 + \hat{c}_{12} ) + \beta_0 ( \partial_1 \hat{c}_{23} + \partial_2 \hat{c}_{31} + \partial_3 \hat{c}_{12} ) \right] \\
      &\quad \quad + \frac{iN}{2\pi} \left[ - c_{12} ( -\partial_3 \hat{C}^2_0 + \partial_0 \hat{C}^2_3 ) - c_{23} ( \partial_0 \hat{C}^3_1 - \partial_1 \hat{C}^3_0 ) - c_{31}( \partial_0 \hat{C}^1_2 - \partial_2 \hat{C}^1_0 ) \right. \\ 
      &\quad \quad \left. - c_{01} ( \partial_2 \hat{C}^1_3 - \partial_1 \hat{C}^1_2 + \hat{c}_{23} ) - c_{02}( \partial_3 \hat{C}^2_1 - \partial_1 \hat{C}^2_3 + \hat{c}_{31} ) - c_{03} ( \partial_1 \hat{C}^3_2 - \partial_2 \hat{C}^3_1 + \hat{c}_{12} )  \right] \\
      &\quad \simeq \frac{iN}{2\pi} \sum_{(i,j,k)} \left[  \beta_{ij} ( \partial_k \hat{C}_0^{k(ij)} - \partial_0 \hat{C}^{ij} )  + \beta_{0k} ( \partial_i \hat{C}^k_j - \partial_j \hat{C}^k_i + \hat{c}_{ij} ) + \beta_0 \partial_k \hat{c}_{ij}   \right] \\
      &\quad \quad + \frac{iN}{2\pi} \left[ - c_{12} ( -\partial_3 \hat{C}^2_0 + \partial_0 \hat{C}^2_3 ) - c_{23} ( \partial_0 \hat{C}^3_1 - \partial_1 \hat{C}^3_0 ) - c_{31}( \partial_0 \hat{C}^1_2 - \partial_2 \hat{C}^1_0 ) \right. \\ 
      &\quad \quad \left. - c_{01} ( \partial_2 \hat{C}^1_3 - \partial_1 \hat{C}^1_2 + \hat{c}_{23} ) - c_{02}( \partial_3 \hat{C}^2_1 - \partial_1 \hat{C}^2_3 + \hat{c}_{31} ) - c_{03} ( \partial_1 \hat{C}^3_2 - \partial_2 \hat{C}^3_1 + \hat{c}_{12} )  \right] \,. 
}
In this formula, we can write
\alis{
    \partial_i \hat{C}^k_j - \partial_j \hat{C}^k_i + \hat{c}_{ij} &= \partial_i ( \hat{C}^k_j - \hat{C}^i_j) + \partial_j (\hat{C}^j_i - \hat{C}^k_i ) +  \partial_i \hat{C}^i_j - \partial_j  \hat{C}^j_i + \hat{c}_{ij} \\
    &\simeq \partial_i \hat{C}^{ki} + \partial_j \hat{C}^{jk} +  \hat{C}^k \,,
}
and
\alis{
    \sum_{(i,j,k)} \partial_k \hat{c}_{ij}  &= \sum_{(i,j,k)} \left[ \partial_k (\partial_i \hat{C}_j^i - \partial_j \hat{C}_i^j + \hat{c}_{ij} ) + \partial_i \partial_j ( \hat{C}^i_k - \hat{C}^j_k ) \right] \\
    &= \sum_{(i,j,k)} ( \partial_k \hat{C}^k + \partial_i \partial_j  \hat{C}^{ij} )  \,,
}
so we derive the $\L^3_{\text{SSPT,e},\beta}$ part
\alis{
    \L^3_{\text{SSPT,e},\beta} &= \frac{iN}{2\pi} \sum_{(i,j,k)} \left[  \beta_{ij} ( \partial_k \hat{C}_0^{k(ij)} - \partial_0 \hat{C}^{ij} ) + \beta_{0k} ( \partial_i \hat{C}^{ki} + \partial_j \hat{C}^{jk} +  \hat{C}^k  ) +  \beta_0 ( \partial_k \hat{C}^k + \partial_i \partial_j  \hat{C}^{ij} )  \right] \\
    & \quad + \frac{iN}{2\pi} \left[ - c_{12} ( -\partial_3 \hat{C}^2_0 + \partial_0 \hat{C}^2_3 ) - c_{23} ( \partial_0 \hat{C}^3_1 - \partial_1 \hat{C}^3_0 ) - c_{31}( \partial_0 \hat{C}^1_2 - \partial_2 \hat{C}^1_0 ) \right. \\ 
    & \quad \left. - c_{01} ( \partial_2 \hat{C}^1_3 - \partial_1 \hat{C}^1_2 + \hat{c}_{23} ) - c_{02}( \partial_3 \hat{C}^2_1 - \partial_1 \hat{C}^2_3 + \hat{c}_{31} ) - c_{03} ( \partial_1 \hat{C}^3_2 - \partial_2 \hat{C}^3_1 + \hat{c}_{12} )  \right] \,. \label{3ssptbeta}
}
Regarding the $\L^3_{\text{SSPT},\text{f}, C\hat{C}}$ part, we can write as
\alis{
    \L^3_{\text{SSPT},\text{f}, C\hat{C}} &= \frac{iN}{2\pi}  \sum_{k=1}^3 \hat{C}^k \wedge  C^k \wedge dx^k + \frac{iN}{2\pi} c\wedge \hat{c}  \\
    &= \frac{iN}{2\pi} \left( C^1_{02} \hat{C}^1_3 - C^1_{03} \hat{C}^1_2 + C^1_{23} \hat{C}^1_0  - C^2_{01} \hat{C}^2_3 + C^2_{03} \hat{C}^2_1   \right. \\
    & \quad \left.  + C^2_{31} \hat{C}^2_0 - C^3_{02} \hat{C}^3_1 + C^3_{01} \hat{C}^3_2 + C^3_{12} \hat{C}^3_0 + c_{01} \hat{c}_{23} + c_{02} \hat{c}_{31} + c_{03}\hat{c}_{12} \right) \\
    &\simeq \frac{iN}{2\pi} \sum_{(i,j,k)} \left[ C_{0ij} ( \hat{C}^i_k - \hat{C}^j_k ) + \frac{1}{3} ( C_{i(jk)} - C_{j(ki)} ) \hat{C}^k_0 \right]  \\
    & \quad +  \frac{iN}{2\pi} \big[ - \partial_2 c_{01} \, \hat{C}^1_3 + ( \partial_3 c_{01} - \partial_0 c_{31} ) \hat{C}^1_2 - ( - \partial_2 c_{31} ) \hat{C}^1_0 + ( \partial_1 c_{02} - \partial_0 c_{12} ) \hat{C}^2_3  \\ 
    & \quad  - \partial_3 c_{02} \, \hat{C}^2_1 - (- \partial_3 c_{12} ) \hat{C}^2_3 + ( \partial_2 c_{03} - \partial_0 c_{23} ) \hat{C}^3_1 - \partial_1 c_{03} \hat{C}^3_2 - (- \partial_1 c_{23}) \hat{C}^3_0  \\ 
    & \quad  +   c_{01} \hat{c}_{23} + c_{02} \hat{c}_{31} + c_{03}\hat{c}_{12} \big] \\
    &\simeq  \frac{iN}{2\pi} \sum_{(i,j,k)} \left[ C_{0ij}  \hat{C}^{ij} - \frac{1}{3} C_{k(ij)} \hat{C}_0^{k(ij)} \right] \\
    & \quad +  \frac{iN}{2\pi} \big[ - \partial_2 c_{01} \, \hat{C}^1_3 + ( \partial_3 c_{01} - \partial_0 c_{31} ) \hat{C}^1_2 - ( - \partial_2 c_{31} ) \hat{C}^1_0 + ( \partial_1 c_{02} - \partial_0 c_{12} ) \hat{C}^2_3  \\ 
    & \quad  - \partial_3 c_{02} \, \hat{C}^2_1 - (- \partial_3 c_{12} ) \hat{C}^2_3 + ( \partial_2 c_{03} - \partial_0 c_{23} ) \hat{C}^3_1 - \partial_1 c_{03} \hat{C}^3_2 - (- \partial_1 c_{23}) \hat{C}^3_0  \\
    & \quad +  c_{01} \hat{c}_{23} + c_{02} \hat{c}_{31} + c_{03}\hat{c}_{12} \big]  \,. \label{3ssptcc}
}
Combining the last terms of \eqref{3ssptbeta} and \eqref{3ssptcc} and dropping the $x^0$-, $x^1$- and $x^3$-derivative terms, we have
\ali{
    \frac{iN}{2\pi} \partial_3 \left( c_{01} \hat{C}^1_2 - c_{02} \hat{C}^2_1 + c_{12} \hat{C}^2_0 \right) \,.
}
As in the case of the SSPT phase with two foliations, we have to add the term \eqref{bdcc} when converting the SSPT phase from the foliated form to the exotic form, and this term cancels out.

After all, we have constructed the exotic SSPT phase with three foliations in 3+1 dimensions:
\alis{
    &\L^3_{\text{SSPT},\text{e}} \left[ C_{0ij}, C_{k(ij)}, \hat{C}_0^{k(ij)}, \hat{C}^{ij}, \hat{C}^k \right]  \\
    &\qquad = \frac{iN}{2\pi} \sum_{(i,j,k)} \hat{\beta}^{k(ij)} \left( \partial_k C_{0ij} - \frac{1}{3} \partial_0 C_{k(ij)} \right) \\ 
    & \qquad \quad + \frac{iN}{2\pi} \sum_{(i,j,k)} \left[  \beta_{ij} ( \partial_k \hat{C}_0^{k(ij)} - \partial_0 \hat{C}^{ij} ) + \beta_{0k} ( \partial_i \hat{C}^{ki} + \partial_j \hat{C}^{jk} +  \hat{C}^k  ) +  \beta_0 ( \partial_k \hat{C}^k + \partial_i \partial_j  \hat{C}^{ij} )  \right] \\
    & \qquad \quad + \frac{iN}{2\pi} \sum_{(i,j,k)} \left[ C_{0ij}  \hat{C}^{ij} - \frac{1}{3} C_{k(ij)} \hat{C}_0^{k(ij)} \right] \,. \label{3foliesspt}
}

As in the other cases, if the theory is on spacetime without a boundary, it is gauge invariant. If spacetime has a boundary, the partition function is not gauge invariant and the variation is canceled by the anomaly of the 2+1d exotic $BF$ theory on the boundary \eqref{eanomaly}. To see this, we put the exotic SSPT phase with three foliations on the region $x^3 \geq 0$ with the boundary $x^3 = 0$. From the gauge invariance, the boundary conditions of $\hat{\beta}^{k(ij)}$, $\beta_{0k}$, $\beta_{ij}$ and $\beta_0$ are
\ali{
    \hat{\beta}^{3(12)}\, |_{x^3 = 0} &= 0 \,,  \\ 
    \beta_{12}\, |_{x^3 = 0} &= 0 \,,  \\
    \beta_0\, |_{x^3 = 0} &= 0  \,.
}
On the boundary, we put the 2+1d exotic $BF$ theory coupled to the background gauge fields \eqref{ebfc}, and the background gauge fields in the 3+1d SSPT phase with three foliations are related to those in the 2+1d exotic $BF$ theory as
\ali{
    C_{\text{SSPT},012}\, |_{x^3 = 0} &= C_{BF,012} \,, \label{3ecbou1} \\
    \hat{C}^{3(12)}_{\text{SSPT},0}\, |_{x^3 = 0} &= \hat{C}^{12}_{BF,0} \,, \label{3ecbou2} \\ 
    \hat{C}^3_{\text{SSPT}}\, |_{x^3 = 0} &= \hat{C}_{BF} \,. \label{3ecbou3}
}
Then, under the background gauge transformations, the Lagrangian transforms as
\alis{
    &\delta_g \L^3_{\text{SSPT,e}} \\
    & \quad =  \frac{iN}{2\pi} \sum_{(i,j,k)} \hat{\Lambda}^{k(ij)} \left( \partial_k C_{0ij} - \frac{1}{3} \partial_0 C_{k(ij)} \right) \\ 
    & \quad \quad + \frac{iN}{2\pi} \sum_{(i,j,k)} \left[  - \Lambda_{ij} ( \partial_k \hat{C}_0^{k(ij)} - \partial_0 \hat{C}^{ij} )  - \partial_k \Lambda_0 ( \partial_i \hat{C}^{ki} + \partial_j \hat{C}^{jk} +  \hat{C}^k  ) -  \Lambda_0 ( \partial_k \hat{C}^k + \partial_i \partial_j  \hat{C}^{ij} )  \right] \\
    & \quad \quad + \frac{iN}{2\pi} \sum_{(i,j,k)} \left[ ( \partial_0 \Lambda_{ij} - \partial_i \partial_j \Lambda_0 ) \hat{C}^{ij}  + ( C_{0ij} + \partial_0 \Lambda_{ij} - \partial_i \partial_j \Lambda_0 ) \partial_k \hat{\Lambda}^{k(ij)} \right] \\
    & \quad \quad -  \frac{iN}{2\pi}   \sum_{(i,j,k)} \frac{1}{3} \left[   ( 2\partial_k \Lambda_{ij} - \partial_i \Lambda_{jk} - \partial_j \Lambda_{ki} ) \hat{C}_0^{k(ij)} +  ( C_{k(ij)} + 2\partial_k \Lambda_{ij} - \partial_i \Lambda_{jk} - \partial_j \Lambda_{ki} ) \partial_0 \hat{\Lambda}^{k(ij)} \right] \\
    & \quad = \frac{iN}{2\pi} \partial_3 \left[ \hat{\Lambda}^{3(12)} C_{012}  -  \Lambda_{12} \hat{C}_0^{3(12)} - \Lambda_0 \hat{C}^3 - \Lambda_{12} \partial_0 \hat{\Lambda}^{3(12)} - \Lambda_0 \partial_1 \partial_2 \hat{\Lambda}^{3(12)} \right] \,,
}
where we have used equations such as
\alis{
    \sum_{(i,j,k)} ( \partial_k \Lambda_{ij} + \partial_i \Lambda_{jk} + \partial_j \Lambda_{ki} ) \hat{C}_0^{k(ij)} &= \sum_{(i,j,k)} \partial_k \Lambda_{ij}  (\hat{C}_0^{k(ij)} + \hat{C}_0^{j(ki)} + \hat{C}_0^{i(jk)} ) \\
    &= 0 \,.
}
Thus, on the boundary $x^3 = 0$, the term
\alis{
    &\delta_g S^3_{\text{SSPT,e}} \\
    & \quad = -\int dx^0 dx^1 dx^2 \frac{iN}{2\pi}  \left[ \hat{\Lambda}^{3(12)} C_{012} -\Lambda_{12} ( \hat{C}^{3(12)}_0 + \partial_0 \hat{\Lambda}^{3(12)} ) -  \Lambda_0 ( \hat{C}^3 + \partial_1 \partial_2 \hat{\Lambda}^{3(12)} )   \right]_{x^3 =0} 
}
arises. From the boundary conditions \eqref{3ecbou1}--\eqref{3ecbou3}, the background gauge parameters also satisfy
\ali{
    \Lambda_{\text{SSPT},0}\, |_{x^3 = 0} &= \Lambda_{BF,0} \,, \\
    \Lambda_{\text{SSPT},12}\, |_{x^3 = 0} &= \Lambda_{BF,12} \,, \\
    \hat{\Lambda}^{3(12)}_{\text{SSPT},0}\, |_{x^3 = 0} &= \hat{\Lambda}^{12}_{BF,0}  
}
on the boundary, and then it matches the 't Hooft anomaly of the 2+1d exotic $BF$ theory \eqref{eanomaly}. Therefore we can also cancel the 't Hooft anomaly of the 2+1d exotic $BF$ theory on the boundary by the gauge-variation of the 3+1d exotic SSPT phase with three foliations on the bulk.

In the foliated form, the SSPT phase with two foliations \eqref{2folisspt4} are related to the one with three foliations \eqref{3foliesspt} in a rather simple way. On the other hand, in the exotic form, relation between the SSPT phase with two foliations \eqref{sspte} and three foliations \eqref{3foliesspt} is non-trivial. To convert the SSPT phase from the foliated form to the exotic form, we must integrate out $\hat{c}_{01}$, $\hat{c}_{02}$, $\hat{c}_{03}$ and $p$. However, in the case of the foliated SSPT phases, they are smoothly deformed into each other preserving the 90 degree rotational symmetry $\Z_4$ provided that the bulk gauge field $\hat{c}$ is not dynamical. Thus we cannot naively deform the exotic SSPT phase with two foliations into the one with three foliations. This is an obscure point of the deformation.

\section{Conclusion}

In this work, we have discussed the mixed 't Hooft anomaly of subsystem symmetry in the exotic and foliated $BF$ theories in 2+1 dimensions and the SSPT phases in 3+1 dimensions that cancel it via the anomaly inflow. We have constructed the exotic and foliated SSPT phases with two and three foliations respectively by using the foliated-exotic duality. Along the way, we have shown the non-topological defect that describes a fracton can be considered as a symmetry-like operator. We have also seen that both of the SSPT phase with two foliations and three foliations match the same 't Hooft anomaly of the exotic/foliated $BF$ theory, and have pointed out that this fact may be a clue for characterizing 't Hooft anomalies of subsystem symmetry.

One of the future directions is to further investigate the anomaly inflow for subsystem symmetries. In this paper, we have considered the 3+1d SSPT phase with two foliations $e^1 =dx^1$ and $e^2 = dx^2$ with the 2+1d exotic/foliated $BF$ theory with two foliations on the boundary $x^3 = 0$. If the boundary is $x^2 = 0$, the boundary theory would be the 2+1d exotic/foliated $BF$ theory with one foliation. Furthermore, it is interesting to put the SSPT phase on the region $x^2 \geq 0$ and $x^3 \geq 0$ with the boundary $x^2 = 0$ and $x^3 = 0$ with a corner. This situation is related to higher-order SSPT phases \cite{You:2019ugx,May-Mann:2022fvm,Zhang:2022wik}, where the anomaly theory would arise on the corners or hinges. Even if the SSPT phase we have considered does not have the corner theory, there may be effective field theories of such higher-order SSPT phases, and finding them is also an interesting topic. These studies will be connected to larger goals, which are the characterization of the 't Hooft anomaly and the classification of SSPT phases.

The other direction is to expand the foliated-exotic duality. There are gapless fractonic theories (e.g., the $\phi$ theory \cite{Seiberg:2020bhn}) in the exotic form, but the corresponding foliated QFTs have not yet been found. Moreover, it is interesting to consider relation between exotic/foliated theories and other topics on fractonic theory, such as the boson-fermion duality with subsystem symmetry \cite{Cao:2022lig}, the infinite-component Chern-Simons-Maxwell theory \cite{Chen:2022hbz,Chen:2023oov}, and the non-invertible duality interfaces with subsystem symmetries \cite{Spieler:2024fby}. Since exotic form and foliated form have different manifest structure, clarifying the correspondences will lead to a deeper understanding of fractonic QFTs and subsystem symmetries.

\newpage

\subsection*{Acknowledgement}
I would like to thank Kantaro Ohmori for his many comments on this work in the discussions. I would also like to thank Yuji Tachikawa for his helpful comments. This work is supported by the World-leading INnovative Graduate Study Program for Frontiers of Mathematical Sciences and Physics (WINGS-FMSP), The University of Tokyo.

\bibliography{ref} 
\bibliographystyle{utphys}

\end{document}